\documentclass[11pt]{article}
\pdfoutput = 1
\usepackage[utf8]{inputenc}

\usepackage[top=1in,bottom=1in,left=1in,right=1in]{geometry}
\makeatletter \g@addto@macro\@floatboxreset\centering \makeatother
\usepackage{color}
\definecolor{darkgreen}{rgb}{0,0.5,0}
\definecolor{darkblue}{rgb}{0,0,0.6}
\definecolor{purple}{rgb}{0.4,.2,0.7}
\usepackage[noEucal]{main}
\usepackage{aas_macros,empheq,fancybox,graphicx,multirow}
\usepackage{aas_macros}
\usepackage{dsfont}
\usepackage{esint}
\usepackage{float}
\usepackage{appendix}
\usepackage{dsfont}
\usepackage{mathabx}
\usepackage{hhline}
\usepackage{vcell}
\usepackage{tensor} 
\usepackage{footmisc}
\usepackage{tabularray}
\usepackage{float}
\usepackage{caption}
\usepackage{relsize}
\DeclareSymbolFont{cmlargesymbols}{OMX}{cmex}{m}{n}
\let\sumop\relax
\DeclareMathSymbol{\sumop}{\mathop}{cmlargesymbols}{"50}

\usepackage[framemethod=TikZ]{mdframed}
\newenvironment{theo}[1][]{
\mdfsetup{
frametitle={
\tikz[baseline=(current bounding box.east),outer sep=0pt]
\node[anchor=east,rectangle,fill=blue!20]
{#1};}}
\mdfsetup{innertopmargin=10pt,linecolor=blue!20,
linewidth=2pt,topline=true,
frametitleaboveskip=\dimexpr-\ht\strutbox\relax,}
\begin{mdframed}[]\relax
}{\end{mdframed}}

\begin{document}

\thispagestyle{empty}

\begin{center}
    ~
    \vskip10mm

     {\LARGE  {\textsc{Supersymmetric Soft Theorems}}}
    \vskip10mm
    
Adam Tropper \\
    \vskip1em
    {\it
        Center for the Fundamental Laws of Nature, \\
Harvard University, Cambridge, Massachusetts 02138, USA\\ \vskip1mm
         \vskip1mm
    }
    \vskip5mm
    \tt{adam$\_$tropper@g.harvard.edu}
\end{center}
\vspace{10mm}

\begin{abstract}
\noindent

We show that in supersymmetric theories, knowing the soft theorem for a single particle in a supermultiplet allows one to immediately determine soft theorems for the remainder of the supermultiplet. While soft theorems in supersymmetric theories have a rich history, they have only been chronicled for specific examples due to the fact that they are usually derived with technical Feynman diagrammatics or amplitudes methods. By contrast, we show that one can compute soft theorems non-perturbatively for entire supermultiplets in one line of algebra. This formalism is directly applicable to the most general supersymmetric theory: one with an arbitrary matter content, number of supercharges, and spacetime dimension. We give many explicit examples illustrating the scope and dexterity of this framework.

\end{abstract}
\pagebreak

\setcounter{tocdepth}{2}
{\hypersetup{linkcolor=black}
\small
\tableofcontents
}

\newpage 

\section{Introduction}

\textit{Soft theorems} relate scattering amplitudes involving extremely low-energy massless particles to those without them. Let $a_i(p_s)$ be the annihilation operator for an outgoing particle of species $i$ and momentum $p_s$. We use crossing symmetry to write amplitudes in an \textit{all-out} formalism. In the \textit{soft limit} $(p_s \rightarrow 0)$, amplitudes involving $a_i(p_s)$ factorize according to a \textit{soft theorem}:
\begin{equation}
    \langle 0| a_{j_1}(p_1) \cdots a_{j_n}(p_n) a_i(p_s) |0\rangle \xrightarrow{p_s \rightarrow 0}  \langle 0| a_{j_1}(p_1) \cdots a_{j_n}(p_n) |0\rangle \bullet \mathcal{S}_{i}(p_s),
    \label{eqn: soft theorem definition}
\end{equation}
where $\mathcal{S}_i(p_s)$ -- the so-called \textit{soft operator} associated to $i$ -- is a linear operator acting on the space of scattering amplitudes. Though this notation is non-standard, it will ultimately prove useful to view $\mathcal{S}_i(p_s)$ as acting from the right on an amplitude; for now, it is just a naming convention.

In the soft limit, $\mathcal{S}_i(p_s)$ may be series expanded in $p_s$. Sometimes the scattering amplitude on the left hand side of Equation \eqref{eqn: soft theorem definition} will diverge in the soft limit -- in this case, the series expansion contains a $1/p_s$ pole. Other times, the amplitude may be finite or even vanish. The information about the entire analytic structure of the amplitude is captured by the coefficients in this series expansion. At times, the first few most-singular terms are known exactly. For example, the leading-order terms when $i$ is a photon, gluon, or graviton are fixed by Lorentz and gauge invariance \cite{Weinberg:1965nx,Low:1958sn}. Such expressions are \textit{universal}, meaning they are theory independent; i.e. any QFT with a $U(1)$ gauge boson must obey the leading soft photon theorem.

It is no surprise that soft theorems and their universality have played a central role the amplitudes program \cite{Cheung:2014dqa, Cheung:2015ota, Cheung:2016drk, Zhou:2022orv, Kalyanapuram:2020epb, Bautista:2019tdr}. Remarkably, their influence is much broader. For example, soft theorems are the Ward identities of infinite-dimensional symmetry algebras known as \textit{asymptotic symmetries} \cite{Strominger:2017zoo,Strominger:2013jfa,Strominger:2013lka, Lysov:2014csa, He:2014laa, He:2014cra, Miller:2021hty}. They also give rise to so-called \textit{memory effects} which have experimental implications \cite{Strominger:2014pwa, Pasterski:2015tva, Pate:2017vwa,Chatterjee:2017zeb, Laddha:2018vbn, Ball:2018prg}. Moreover, they play a central role in \textit{celestial holography}, as they encode conserved currents on the putative CFT living on the conformal boundary of asymptotically flat spacetime \cite{Raclariu:2021zjz,Pasterski:2021rjz,Pasterski:2023ikd, Kapec:2016jld, Nande:2017dba, Himwich:2020rro, Strominger:2021lvk, Donnay:2022hkf}. Due to their wide-ranging relevance, an interest in developing tools to study new soft theorems in novel circumstances has emerged \cite{Cachazo:2014fwa, Strominger:2015bla, Hawking:2016msc, Hawking:2016sgy,Fernandes:2020tsq, Marotta:2019cip, Miller:2022fvc, Tropper:2023fjr, Herderschee:2023bnc, Godazgar:2019dkh, Choi:2019rlz, Laddha:2017ygw, Cheung:2021yog, Cheung:2023qwn, Cheng:2022xyr, Cheng:2022xgm, Assassi:2012zq, Derda:2024jvo}. The present work is consistent with that ethos. 

We focus on soft theorems in supersymmetric theories and how they are related to one another; henceforth, \textit{supersymmetric soft theorems}.

\textbf{Central question:} Certain soft theorems are universal or otherwise known explicitly. Can the constraints imposed by supersymmetry be leveraged to yield soft theorems associated to an arbitrary particle in the same supermultiplet as the particle whose soft theorem is known?

This question was originally answered in the affirmative in \cite{Dumitrescu:2015fej} in the specific case of $d = 4$, $\mathcal{N} = 1$ supersymmetric QED. While the result was suggestive, the derivation could not be easily extended beyond this limited regime. Since then, studying soft theorems in supersymmetric theories has gained a great deal of interest, particularly in the celestial holography program \cite{Liu:2014vva, Lysov:2015jrs, Avery:2015iix, Chen:2014xoa, Bork:2015fla, Jain:2018fda, Rao:2014zaa, Pano:2021ewd, Fotopoulos:2020bqj, Ball:2023qim, Bu:2021avc, Jiang:2021ovh, Jiang:2021xzy, Banerjee:2022hpo}. A notable drawback is that these works typically have a very specific QFT in mind, and the task of determining soft theorems for the supermultiplet often boils down to specifying a Lagrangian and then computing soft theorems with very sophisticated, technical, and arduous Feynman diagrammatics. In this way, such results -- while extremely important to studying certain interesting models -- are neither generalizable nor fully harness the immense power of supersymmetry. In this work, we demonstrate that there is a much cleaner route.

\textbf{Central finding:} Using the supersymmetric Ward identity and basic representation theory, one may leverage known soft theorems for some particle to completely determine the soft theorems for that particle's entire supermultiplet. The analysis is fully non-perturbative, and the computations boil down a single line of algebra. This idea is also immediately applicable to the most general supersymmetric theory one can imagine: one with arbitrarily many supercharges, arbitrary matter content, and in an arbitrary number of dimensions.

\textbf{Outline:} The body of this article is in $d = 4$ spacetime dimensions due to the simple fact that the supersymmetry algebra and spinor representations depend intricately on this choice. It is pedagogically simpler to just pick $d=4$ for the moment, rather than immediately dealing with casework. In Section \ref{sec: SUSY soft theorems}, we review the basics of supersymmetry and present our central finding about how to determine supersymmetric soft theorems. In Section \ref{sec: examples}, we provide specific examples in supersymmetric gauge and gravitational theories. In Appendix \ref{appendix: on shell superspace}, we show that our main results take an especially elegant form when written in the language of \textit{on-shell superspace}. In Appendix \ref{appendix: arbitrary dimensions}, we systematically re-present our findings in a manner which is consistent with working in an arbitrary number of dimensions.

\section{Supersymmetric Soft Theorems}
\label{sec: SUSY soft theorems}

In this section, we discuss how to leverage the soft theorem for one particle in a supermultiplet to determine the soft theorems for the rest of the supermultiplet. We begin by reviewing the basics.

\subsection{Massless Supersymmetry Multiplets}

In four dimensions, supersymmetric theories are characterized by $\mathcal{N}$ supercharges which transform as two-component Weyl spinors. These supercharges are labeled $Q_\alpha^{I}$ and $\overline{Q}^I_{\dot{\alpha}} = (Q^I_{\alpha})^\dagger$ where $I = 1, ..., \hspace{2pt}\mathcal{N}$ and $\alpha = 1,2$. The super-Poincar\'e algebra consists of the usual Poincar\'e algebra supplemented by the following commutation relations between supercharges and the Poincar\'e generators\footnote{Following \cite{Srednicki:2007qs, Elvang:2013cua}, we use the mostly-plus metric convention. Pauli matrices are $\sigma^{\mu}_{\alpha \dot{\alpha}} = (\mathbf{1}_{\alpha \dot{\alpha}},\sigma^i_{\alpha \dot{\alpha}})$ and $\overline{\sigma}^{\mu \dot{\alpha} \alpha} = \varepsilon^{\dot{\alpha} \dot{\beta}} \varepsilon^{\alpha \beta} \sigma^{\mu}_{\beta \dot{\beta}} = (\mathbf{1}^{\dot{\alpha} \alpha},-\sigma^{i,\dot{\alpha} \alpha})$. Lorentz generators in the spinor representation are $\sigma^{\mu \nu} = \tfrac{i}{2}\sigma^{[\mu}\overline{\sigma}^{\nu]}$ and $\overline{\sigma}^{\mu \nu} = -\tfrac{i}{2}\overline{\sigma}^{[\mu}\sigma^{\nu]}$.}
\begin{equation}
    \begin{split}
        \Big[J^{\mu \nu},Q_\alpha^I\Big] &= -(\sigma^{\mu \nu})_\alpha^{~ \beta}\hspace{2pt} Q_\beta^I \hspace{70pt} \Big[P_\mu,Q_\alpha^I\Big] = 0 \\
        \Big[J^{\mu \nu},\overline{Q}^{I}_{\dot{\alpha}}\Big] & =-(\overline{\sigma}^{\mu \nu})_{\dot{\alpha}}^{~ \dot{\beta}} \hspace{2pt} \overline{Q}^{I}_{ \dot{\beta}} \hspace{70pt} \Big[P_\mu,\overline{Q}^{I}_{ \dot{\alpha}}\Big] = 0, \\
    \end{split}
\end{equation}
as well as between the supercharges and themselves
\begin{equation}
    \Big[Q_\alpha^I,\overline{Q}_{\dot{\alpha}}^J\Big] = -2 \sigma^{\mu}_{\alpha \dot{\alpha}} P_\mu \delta^{IJ} \hspace{50pt} \Big[Q_\alpha^I,Q_\beta^J\Big] = \varepsilon_{\alpha \beta}\hspace{1pt} Z^{IJ} \hspace{50pt} \Big[\hspace{2pt}\overline{Q}_{\dot{\alpha}}^I,\overline{Q}_{\dot{\beta}}^J\Big] = \varepsilon_{\dot{\alpha}\dot{\beta}} \hspace{1pt} \overline{Z}^{IJ}.
\end{equation}
Here $Z^{IJ}$ is an antisymmetric matrix of central charges, and $\overline{Z}^{IJ} = (Z^{IJ})^*$. Note that the supersymmetry algebra is a $\mathbb{Z}_2$-graded Lie algebra where the Poincar\'e generators are bosonic and the supersymmetry generators are fermionic. The bracket $\big[\cdot,\cdot\big]$ will always signify the \textit{Lie superbracket}, which becomes the standard commutator or anticommutator according to the parity of its entries. We will denote the parity of an operator $\mathcal{O}$ by
\begin{equation}
    |\mathcal{O}| = \begin{cases}
        ~ ~ 0 \hspace{30pt} \mathcal{O} ~ ~ \text{is bosonic}\\
        ~ ~ 1 \hspace{30pt} \mathcal{O} ~ ~ \text{is fermionic.}\\
    \end{cases}
\end{equation}

Supersymmetry multiplets may be constructed by starting with a highest-weight state of the supersymmetry algebra. This state is annihilated by some operator, which we call $a(p).$ We focus on massless states because only these particles enjoy soft theorems in the conventional sense. The remainder of the supersymmetry multiplet is obtained by acting on $a(p)$ with all possible strings of supersymmetry generators subject to the supersymmetry algebra and the highest-weight condition $\big[a(p),\overline{Q}^I_{\dot{\alpha}}\big] = 0$. This defines the annihilation operators $a^{I_1 \cdots I_n}(p)$ with $0 \leq n \leq \mathcal{N}$ via
\begin{equation}
    \Big[a^{I_1 \cdots I_n}(p),Q^{I}_{\alpha}\Big] = \sqrt{2} \hspace{2pt} |p]_{\alpha} \hspace{2pt} a^{I I_1 \cdots I_n}(p) \hspace{40pt} \Big[a^{I_1 \cdots I_n}(p),\overline{Q}^I_{\dot{\alpha}}\Big] = \sqrt{2} \hspace{2pt} \langle p|_{\dot \alpha} \hspace{2pt} n \hspace{2pt} \delta^{I[I_1} a^{I_2 \cdots I_n]}(p),
    \label{eqn: SUSY multiplet}
\end{equation}
where $\langle p|_{\dot{\alpha}}$ and $|p]_{\alpha}$ are spinor-helicity variables satisfying $p_\mu \sigma^{\mu}_{\alpha \dot{\alpha}} = -|p]_\alpha \langle p|_{\dot{\alpha}}.$
Together, these $a^{I_1 \cdots I_n}(p)$ constitute the supermultiplet of the highest-weight state $a(p)$. If this representation is not CPT self-conjugate, one must also include a second supermultiplet consisting of the opposite helicity CPT conjugates $\overline{a}^{I_1 \cdots I_n}(p)$. Such operators satisfy:
\begin{equation}
    \Big[\hspace{2pt} \overline{a}^{I_1 \cdots I_n}(p),\overline{Q}^I_{\dot \alpha}\Big] = \sqrt{2}\hspace{2pt} \langle p|_{\dot \alpha} \hspace{2pt} \overline{a}^{I I_1 \cdots I_n}(p) \hspace{40pt} \Big[\hspace{2pt} \overline{a}^{I_1 \cdots I_n}(p),Q^I_{\alpha}\Big] = \sqrt{2}\hspace{2pt} 
    |p]_{\alpha} \hspace{2pt} n \hspace{2pt} \delta^{I[I_1} \overline{a}^{I_2 \cdots I_n]}(p).
    \label{eqn: conjugate SUSY multiplet}
\end{equation}
Note that $\overline{a}(p)$ is a lowest-weight state in the sense that $\big[\overline{a}(p),Q^I_{\alpha}\big] = 0$.

For example, in $\mathcal{N} = 1$ supersymmetric QED, $a(p)$ and $\overline{a}(p)$ are respectively positive and negative helicity photons, while $a^1(p) ~ \propto ~ [a(p),Q^1]$ and $\overline{a}^1 ~ \propto ~ [\overline{a}(p),\overline{Q}^1]$ are respectively positive and negative helicity photinos. Generally, if $a(p)$ has helicity $h$, then $\overline{a}(p)$ has helicity $-h$, while $a^{I_1 \cdots I_n}(p)$ and $\overline{a}^{I_1 \cdots I_n}(p)$ have helicity $h - \tfrac{n}{2}$ and $-h + \tfrac{n}{2}$ respectively. The annihilation operators $a^{I_1 \cdots I_n}(p)$ are totally antisymmetric in their indices, and each additional index swaps their statistics between boson and fermion. Finally, If the highest-weight states are canonically normalized by $[a(p),a^\dagger(p')] = 2E_p \hspace{2pt} \delta^{(3)}(\vec{p} - \vec{p}\hspace{2pt}')$, then all other states in the supermultiplet are also canonically normalized
\begin{equation}
    \big[a^{I_1 \cdots I_n}(p),a^{J_1 \cdots J_m \dagger}(p')\big] = 2E_p \hspace{2pt} \delta^{I_1 \cdots I_n}_{J_1 \cdots J_m} \delta^{(3)}(\vec{p} - \vec{p}\hspace{2pt}'),
\end{equation} 
where $\delta^{I_1 \cdots I_n}_{J_1 \cdots J_m}$ enforces that indices are the same up to totally antisymmetric permutations.

\subsection{Supersymmetry Constraints on Amplitudes}

Henceforth, we distinguish the annihilation operators $a^{I_1 \cdots I_n}(p)$ introduced previously with the annihilation operator $a_j(p)$ encoding some arbitrary particle $j$ in the theory. Of course, $a_j(p)$ lives in some supersymmetry multiplet of its own; hence, it may always be written in the above form if it is massless (and in a similar manner if it is massive). We also write scattering amplitudes in an \textit{all-out formalism} with $a_j(p)$ acting on the out-state vacuum 
\begin{equation}
    \mathcal{A}_n = \langle 0| a_{j_1}(p_1) \cdots a_{j_n}(p_n)|0\rangle.
\end{equation} 

In this work, we will often have various operators, $\mathcal{O}$, ``acting'' on scattering amplitudes, written $\mathcal{A}_n \bullet \mathcal{O}$. They act via commutator with the asymptotic particles, one-by-one:
\begin{equation}
    \langle 0| a_{j_1}(p_1) \cdots a_{j_n}(p_n) |0\rangle \hspace{1pt}\bullet \hspace{1pt} \mathcal{O}  = \sumop_{m=1}^n (-1)^{\hspace{2pt}\sumop\limits_{m'=m+1}^n \hspace{-9pt}|\mathcal{O}||j_{m'}|} ~ \langle 0 | a_{j_1}(p_1) \cdots \Big[a_{j_m}(p_m),\mathcal{O}\Big]\cdots a_{j_n}(p_n) |0\rangle.
\end{equation}
It is crucial to write this as $\mathcal{A}_n \bullet \mathcal{O}$ rather than $\mathcal{O} \bullet \mathcal{A}_n$ because $\mathcal{O}$ acts by commutator from the \textit{right}. This ordering is also essential for keeping track of minus signs in supercommutators.

When $\mathcal{O}$ is a symmetry generator, it is equipped with the conservation law: $\mathcal{A}_n \bullet \mathcal{O} = 0$. This follows from noting that $\mathcal{O}|0\rangle = 0$ (assuming the vacuum enjoys the symmetry) and then repeatedly (anti-)commuting $\mathcal{O}$ through $\langle 0|a_{j_1}(p_1) \cdots a_{j_n}(p_n) \mathcal{O}|0\rangle = 0$ until it annihilates the out state vacuum. The conservation laws $\mathcal{A}_n \bullet Q^I_{\alpha} = 0 = \mathcal{A}_n \bullet \overline{Q}^I_{\dot{\alpha}} $ are known as \textit{supersymmetric Ward identities}
\begin{equation}
    \begin{split}
        \sumop_{m=1}^n (-1)^{\hspace{2pt}\sumop\limits_{m'=m+1}^n \hspace{-9pt}|j_{m'}|} ~ \langle 0 | a_{j_1}(p_1) \cdots \big[a_{j_m}(p_m),Q^I_{\alpha}\big]\cdots a_{j_n}(p_n) |0\rangle &= 0 \\
        \sumop_{m=1}^n (-1)^{\hspace{2pt}\sumop\limits_{m'=m+1}^n \hspace{-9pt}|j_{m'}|} ~ \langle 0 | a_{j_1}(p_1) \cdots \big[a_{j_m}(p_m),\overline{Q}^I_{\dot{\alpha}}\big]\cdots a_{j_n}(p_n) |0\rangle &= 0.
    \end{split}
\end{equation}

\subsection{Derivation of Supersymmetric Soft Theorems}

We are now ready to derive the supersymmetric soft theorems. Say we know that $\mathcal{S}(p_s)$ is the soft operator associated to some highest-weight state $a(p_s)$ in a supersymmetry multiplet. One may determine the soft theorem for the descendant $a^I(p_s)$ from the supersymmetric Ward identity:
\begin{align}
        \langle 0|a_{j_1}(p_1)& \cdots a_{j_n}(p_n) \hspace{2pt} a^I(p_s)|0\rangle \nonumber \\
        &= \frac{[\xi|^{\alpha}}{\sqrt{2}[\xi \hspace{1pt} p_s]} \langle 0| a_{j_1}(p_1) \cdots a_{j_n}(p_n) \hspace{2pt} \big[a(p_s),Q^I_{\alpha}\big]|0\rangle \nonumber \\
        &= - (-1)^{|a|} \frac{[\xi|^{\alpha}}{\sqrt{2}[\xi\hspace{1pt} p_s]} ~ \sumop_{m=1}^n (-1)^{\hspace{2pt}\sumop\limits_{m'=m+1}^n \hspace{-9pt}|j_{m'}|} ~ \langle 0| a_{j_1}(p_1) \cdots \big[a_{j_m},Q^I_{\alpha}\big] \cdots a_{j_n}(p_n) \hspace{1pt} a(p_s) |0\rangle \nonumber\\
        & \xrightarrow{p_s \rightarrow 0} - (-1)^{|a|} \frac{[\xi|^{\alpha}}{\sqrt{2}[\xi \hspace{1pt} p_s]} ~ \sumop_{m=1}^n (-1)^{\hspace{2pt}\sumop\limits_{m'=m+1}^n \hspace{-9pt}|j_{m'}|} ~ \langle 0|a_{j_1}(p_1) \cdots \big[a_{j_m},Q^I_{\alpha}\big] \cdots a_{j_n}(p_n) |0\rangle \nonumber \bullet \mathcal{S}(p_s) \\
        & \xrightarrow{p_s \rightarrow 0} - (-1)^{|a|} \frac{[\xi|^{\alpha}}{\sqrt{2}[\xi \hspace{1pt} p_s]} ~ \langle 0|a_{j_1}(p_1) \cdots a_{j_n}(p_n) |0\rangle \nonumber \bullet Q^I_{\alpha} \bullet \mathcal{S}(p_s) \\
        & \xrightarrow{p_s \rightarrow 0} \frac{[\xi|^{\alpha}}{\sqrt{2}[\xi \hspace{1pt} p_s]} ~ \langle 0|a_{j_1}(p_1) \cdots a_{j_n}(p_n) |0\rangle \bullet \Big[\mathcal{S}(p_s), Q^I_{\alpha}\Big].
\end{align}

The first line of this derivation comes from the definition of $a^I(p_s)$ as a descendant of $a(p_s)$. The second line is an immediate application of the supersymmetric Ward identity. The third line follows from taking the soft limit. Next, we recognize the summation as just being the operator $Q^I_{\alpha}$ acting on the scattering amplitude. Crucially, $Q^I_{\alpha}$ acts on the scattering amplitude \textit{before} the soft term $\mathcal{S}(p_s)$ does. One might be worried that the fourth line of this expression vanishes due to the supersymmetric Ward identity. This is not the case, however, because $\mathcal{S}(p_s)$ is not just a number -- it is an operator. Indeed, the supersymmetric Ward identity is a sum of terms which delicately cancel; once one starts applying additional operators to these terms individually, they may no longer cancel. Were we to pull $Q^I_{\alpha}$ to the very outside and act with it last, then the expression would vanish by the supersymmetric Ward identity. One can directly check this using the super Jacobi-identity and the fact that the vacuum preserves SUSY, i.e. $Q^I_\alpha|0\rangle = \langle 0|Q^I_{\alpha} = 0.$ The action of ``pulling $Q^I_{\alpha}$ to the outside'' is equivalent to taking its commutator with $\mathcal{S}(p_s)$, where this supercommutator is formally viewed as a commutator of linear operators acting on scattering amplitudes. 

By repeatedly preforming this procedure, one can determine the soft theorem $\mathcal{S}^{I_1 \cdots I_n}(p_s)$ for an arbitrary particle $a^{I_1 \cdots I_n}(p_s)$ in the supersmultiplet (and similarly for the CPT conjugates). Each additional index on the soft operator swaps its statistics between bosonic and fermionic. Such logic yields our main result.

\begin{theo}[Main Result]
Soft theorems for superpartners form a representation of the supersymmetry algebra:
\begin{equation}
    \begin{split}
        \hspace{-20pt}\Big[\mathcal{S}^{I_1 \cdots I_n}(p),Q^{I}_{\alpha}\Big] &= \sqrt{2} \hspace{2pt} |p]_{\alpha} \hspace{2pt} \mathcal{S}^{I I_1 \cdots I_n}(p) \hspace{20pt} \Big[\mathcal{S}^{I_1 \cdots I_n}(p),\overline{Q}^I_{\dot \alpha}\Big] = \sqrt{2} \hspace{2pt} \langle p|_{\dot \alpha} \hspace{2pt} n \hspace{2pt} \delta^{I[I_1} \mathcal{S}^{I_2 \cdots I_n]}(p), \\
        \Big[\overline{\mathcal{S}}^{I_1 \cdots I_n}(p),\overline{Q}^{I}_{\dot \alpha}\Big] &= \sqrt{2} \hspace{2pt} \langle p|_{\dot \alpha} \hspace{2pt} \overline{\mathcal{S}}^{I I_1 \cdots I_n}(p) \hspace{19.5pt} \Big[\overline{\mathcal{S}}^{I_1 \cdots I_n}(p),Q^I_{\alpha}\Big] = \sqrt{2} \hspace{2pt} |p]_{\alpha} \hspace{2pt} n \hspace{2pt} \delta^{I[I_1} \overline{\mathcal{S}}^{I_2 \cdots I_n]}(p).
        \label{eqn: soft theorem SUSY representation}
    \end{split}
\end{equation}
Notice the similarity with Equations \eqref{eqn: SUSY multiplet} and \eqref{eqn: conjugate SUSY multiplet}.
\end{theo}

This may seem like a trivial identity written abstractly in terms of seemingly mysterious commutators between known soft theorems and a supercharge. Nevertheless, this formalism lends itself nicely to myriad examples. In fact, its simplicity is its power. In practice, these commutators can be computed in a single line of algebra. No Feynman diagrammatics are needed, and the significant constraints that supersymmetry enforces on soft theorems become manifest. 

These expressions take an especially elegant form when written in the language of \textit{on-shell superspace}. Nevertheless, this discussion is inessential to our main goal of leveraging knowledge of the soft theorem for a single particle to compute soft theorems for the remainder of its supermultiplet by way of evaluating the commutators of Equation \eqref{eqn: soft theorem SUSY representation}. The details are left to Appendix \ref{appendix: on shell superspace}.

\section{Supersymmetric Soft Theorems in Gauge Theory and Gravity}
\label{sec: examples}

The aim of this section is to illustrate -- by way of several examples -- that computing the commutator $\big[\mathcal{S}^{I_1 \cdots I_n}(p_s),Q^I_\alpha \big]$ is a simple thing to do in practice. We begin by reviewing some foundational aspects of soft theorems and recounting universal expressions for the soft photon and graviton theorems. Afterwards, we compute these commutators explicitly, determining the leading soft theorems for the entire photon and graviton supermultiplets.

\subsection{Soft Theorems for Photons and Gravitons}

It is conventional to parameterize the momentum $p^\mu$ of a massless particle with an overall energy scale $\omega$ and a pair of complex numbers $(z,\bar{z})$ satisfying $z^* = \bar{z}$ for real momentum
\begin{equation}
    p^\mu = \omega \hspace{1pt} \widehat{p}^{\hspace{2pt}\mu} = \omega(1 + z \bar{z},z+\bar{z},-i(z-\bar{z}),1-z\bar{z}).
    \label{eqn: momentum parameterization}
\end{equation}
It is often convenient to series expand the soft operator $\mathcal{S}_i(p_s)$ in the small (compared with the energy scale of the other particles) parameter $\omega_s$\footnote{Equation \eqref{eqn: soft expansion} is only valid at tree level. At loop level, the expansion is corrected by terms of the form $\omega_s^k (\log \omega_s)^l$ \cite{Krishna:2023fxg, Laddha:2018myi, Sahoo:2018lxl, Addazi:2019mjh, Agrawal:2023zea, Ciafaloni:2018uwe, Mao:2023rca}. The formalism established in Section \ref{sec: SUSY soft theorems} can still be applied at loop order, however. Because explicit expressions for loop-corrected soft theorems are extremely complicated and hard to come by in the first place, we will stick to tree-level soft theorems obeying the above expansion for the sake of writing down examples.}
\begin{equation}
    \mathcal{S}_i(p_s) = \sumop_{k \hspace{2pt} \in \hspace{2pt} \mathbb{Z} + \tfrac{1}{2}|i|} \omega^k_s ~ \mathcal{S}^{(k)}_i(\widehat{p}_s).
    \label{eqn: soft expansion}
\end{equation}
The sum ranges over the integers when $i$ is bosonic and half integers when $i$ is fermionic.\footnote{Let us quickly supply a proof of this statement. Under the little group rescaling $|p]_{\alpha} \mapsto t^{-1} |p]_{\alpha}$ and $\langle p|_{\dot \alpha} \mapsto t \langle p|_{\dot \alpha}$, the $n$-particle scattering amplitude involving a particle with momentum $p$ and helicity $h$ gets rescaled as $\mathcal{A}_n \mapsto t^{-2h} \mathcal{A}_n.$ By spin-statistics, $2h$ is even for a bosonic particle and odd for a fermionic one. The spinor brackets for $p$ and $\widehat{p}$ are related by $|p]_{\alpha} = \sqrt{2\omega} \hspace{2pt} |\hspace{1pt} \widehat{p}\hspace{1pt}]_{\alpha}$ and $\langle p|_{\dot \alpha} = \sqrt{2\omega} \hspace{2pt} \langle \hspace{1pt} \widehat{p}\hspace{1pt}|_{\dot \alpha}$ with the equivalent little group rescaling: $\sqrt{\omega} \mapsto t \sqrt{\omega}$, $|\hspace{1pt} \widehat{p}\hspace{1pt}]_{\alpha} \mapsto t^{-2} |\hspace{1pt}\widehat{p}\hspace{1pt}]_{\alpha}$, and $\langle \hspace{1pt}\widehat{p} \hspace{1pt} |_{\dot \alpha} \mapsto \langle \hspace{1pt} \widehat{p} \hspace{1pt} |_{\dot \alpha}.$ For the soft theorem to be consistent, a particular term in the soft expansion $\omega_s^k \mathcal{S}_i^{(k)}(\widehat{p}_s)$ must have the same little group scaling as the particle generating it. However, $\mathcal{S}^{(k)}_i(\widehat{p}_s)$ is some rational function in $|\widehat{p}_s]_{\alpha}, \langle \widehat{p}_s |_{\dot \alpha}$ and the various other external kinematics in the problem. Thus, its little group scaling is $t^{-2m}$ with $2m$ an even number. The parity of the little group scaling exponent, therefore, is completely determined by the power $\omega_s$ is raised to -- an integer power contributes an even amount to the little group scaling exponent and a half-integer power contributes an odd amount.} The most singular term in the $\omega_s$ expansion is called the \textit{leading soft theorem}. Remaining terms are called \textit{subleading soft theorems}.

It follows from Equation \eqref{eqn: soft theorem SUSY representation} that if $a(p_s)$ is a highest-weight state with helicity $h$ and soft theorem $\mathcal{S}(p_s)$, then the soft theorems for $a^{I_1 \cdots I_n}(p_s)$ and the CPT conjugates $\overline{a}^{I_1 \cdots I_n}(p_s)$ are
\begin{equation}
    \begin{split}
        \mathcal{S}^{I_1 \cdots I_n}(p_s) &= \bigg(\prod_{m=1}^n \frac{1}{\sqrt{2}}\frac{[\xi_m|^{\alpha_m}}{[\xi_m \hspace{1pt} \widehat{p}_s]} \bigg) \hspace{2pt} \sumop_{k \hspace{2pt} \in \hspace{2pt} \mathbb{Z} + \tfrac{h}{2}}\omega_s^{k-n/2} \hspace{2pt} \Big[\Big[\Big[ \mathcal{S}^{(k)}(\widehat{p}_s),Q^{I_n}_{\alpha_n}\Big]\hspace{2pt},\cdots\Big]\hspace{2pt},Q^{I_1}_{\alpha_1}\Big] \\
        \overline{\mathcal{S}}^{I_1 \cdots I_n}(p_s) &=  \bigg(\prod_{m=1}^n \frac{1}{\sqrt{2}}\frac{|\xi_m\rangle ^{\dot{\alpha}_m}}{\langle \widehat{p}_s\hspace{1pt} \xi_m \rangle} \bigg) \sumop_{k \hspace{2pt} \in \hspace{2pt} \mathbb{Z} - \tfrac{h}{2}}\omega_s^{k-n/2} \hspace{3pt} \Big[\Big[\Big[\overline{\mathcal{S}}^{(k)}(\widehat{p}_s),\overline{Q}^{I_n}_{\dot{\alpha}_n}\Big]\hspace{2pt},\cdots\Big]\hspace{2pt},\overline{Q}^{I_1}_{\dot \alpha_1}\Big],
        \label{eqn: soft theorem supermultiplet}
    \end{split}
\end{equation}
\vspace{-20pt}

\noindent where $[\xi_m|^{\alpha_m}$ and $|\xi_m\rangle ^{\dot{\alpha}_m}$ are arbitrary reference spinors.

Soft theorems have been especially carefully studied in gauge and gravitational theories. In fact, the leading soft photon theorem is universal and admits a simple expression. Moreover, the first subleading soft photon theorem is also known precisely; however, it is not universal. Indeed, if a certain set of dimension-five operators appear in the effective Lagrangian for the theory, the subleading soft photon theorem will be corrected by a term proportional to the coupling constant of the corresponding dimension-five operator. Explicit expressions for the soft theorems when acting on an $n$-particle scattering amplitude are \cite{Weinberg:1965nx, Low:1958sn, Elvang:2016qvq}
\begin{equation}
    \begin{split}
        \mathcal{S}^{(-1)}_{\text{photon},+}(\widehat{p}_s) &= e \sumop_{m=1}^n q_m \frac{\varepsilon_+(\widehat{p}_s) \cdot p_m}{\widehat{p}_s \cdot p_m} \\
        \mathcal{S}^{(0)}_{\text{photon},+}(\widehat{p}_s) &= - i e \sumop_{m=1}^n q_m \frac{\widehat{p}_{s} \cdot J_m \cdot \varepsilon_+(\widehat{p_s})}{\widehat{p}_s \cdot p_m} + g_k \sumop_{m=1}^n \hspace{2pt} \frac{[\hspace{2pt} \widehat{p}_s \hspace{2pt} p_m]}{\langle\hspace{2pt} \widehat{p}_s \hspace{2pt} p_m\rangle} \hspace{2pt} \mathcal{F}_{k,m},
        \label{eqn: soft photon theorem}
    \end{split}
\end{equation}
Soft theorems for negative helicity photons are similar, being related by \textit{CPT conjugation.}

Here $q_m$ is the charge of the $m^{th}$ particle appearing in the amplitude, $J^{\mu \nu}_m$ is the angular momentum operator acting on the $m^{th}$ particle, and $\varepsilon^\mu_+(\widehat{p}_s)$ is the polarization vector of the positive helicity soft photon. The index $k$ in the final term labels the aforementioned set of dimension-five operators in one's Lagrangian and is implicitly summed over when there are multiple such operators present. $g_{k}$ is the coupling constant of these operators, and $\mathcal{F}_{k,m}$ is a so-called \textit{particle changing operator} which acts on the $m^{th}$ particle. Though the nature of $\mathcal{F}_{k,m}$ depends on the corresponding dimension-five operator, they are known explicitly.\footnote{\label{fnL particle changing}To illustrate that $\mathcal{F}_{k}$ is not a scary object to give one pause, let us briefly summarize its nature in one example \cite{Laddha:2017vfh}. Let $F_{\mu \nu}$ be the photon's field strength, which may be decomposed into self-dual and anti-self dual components $F^\pm_{\mu \nu} = \tfrac{1}{2}(F_{\mu \nu} \mp i \star F_{\mu \nu})$, and let $\varphi$ be a real scalar. Consider the operator $\mathcal{L} \supset g \varphi (F^+_{\mu \nu})^2 + \overline{g} \varphi (F^-_{\mu \nu})^2$. For a positive helicity soft photon, the parameter $g_k$ appearing in Equation \eqref{eqn: soft photon theorem} is just the combination of coupling constants $2e^2g$, and $\mathcal{F}$ has the following action on the various annihilation operators
\begin{equation}
    [a_{\text{scalar}}(p),\mathcal{F}] = a_{\text{photon},-}(p) \hspace{30pt} [a_{\text{photon,+}}(p),\mathcal{F}] = -a_{\text{scalar}}(p) \hspace{30pt} [a_{\text{photon},-}(p),\mathcal{F}] = 0.
\end{equation}
The commutators $[\mathcal{F},Q]$ can now be evaluated with ease using the super-Jacobi identity $[a_j(p),[\mathcal{F},Q]] = [[a_j(p),\mathcal{F}],Q] -[[a_j(p),Q],\mathcal{F}]$ and known relations about how $Q$ and $\mathcal{F}$ move one around a supermultiplet. The $\mathcal{F}$ operators appearing in other theories behave in a totally similar fashion.
}

Similar soft theorems are also understood in non-abelian gauge theory. The electromagnetic charge $e$ gets replaced with $ig_{\text{YM}}$, and there is an extra color charge generator $T^a$ where $a$ matches the color of the soft gluon.

In gravity, the leading and subleading soft theorems are universal and admit simple expressions. The subsubleading soft graviton theorem, however, receives theory dependent corrections proportional to the coupling constants of a certain set of dimension-seven operators which may appear in the effective Lagrangian. Explicit expressions are \cite{Weinberg:1965nx, Cachazo:2014fwa, Elvang:2016qvq}.
\begin{equation}
    \begin{split}
        \mathcal{S}^{(-1)}_{\text{graviton},+}(\widehat{p}_s) &= \frac{\kappa}{2} \sumop_{m=1}^n \frac{\big(p_m \cdot \varepsilon_+(\widehat{p}_s)\big)^2}{\widehat{p}_s \cdot p_m}\\
        \mathcal{S}^{(0)}_{\text{graviton},+}(\widehat{p}_s) &= -\frac{i\kappa}{2} \sumop_{m=1}^n \frac{\big(p_m \cdot \varepsilon_+(\widehat{p}_s)\big) \big(\widehat{p}_s \cdot J_m \cdot \varepsilon_+(\widehat{p}_s)\big)}{\widehat{p}_s \cdot p_m}\\
        \mathcal{S}^{(1)}_{\text{graviton},+}(\widehat{p}_s) &= -\frac{\kappa}{4} \sumop_{m=1}^n \frac{\big(\widehat{p}_s \cdot J_m \cdot \varepsilon_+(\widehat{p}_s)\big)^2}{\widehat{p}_s \cdot p_m} + g_k \sumop_{m=1}^n \frac{[\hspace{2pt} \widehat{p}_s \hspace{2pt} p_m]^3}{\langle\hspace{2pt} \widehat{p}_s \hspace{2pt} p_m\rangle} \hspace{2pt} \mathcal{F}_{k,m},
        \label{eqn: soft graviton theorems}
    \end{split}
\end{equation}
where $\kappa$ is related to Newton's constant by $\kappa = \sqrt{32\pi G_{\text{N}}}$.

\subsection{Evaluating the $[\mathcal{S},Q]$ Commutators}

We begin with the gauge theory case. The leading soft photon theorem of Equation \eqref{eqn: soft photon theorem} is comprised of the following ingredients: some fixed kinematics $\widehat{p}_s^{\hspace{2pt}\mu}$ and $\varepsilon^\mu_+(\widehat{p}_s)$, the momentum operator $P^\mu$ which has eigenvalue $p^\mu_m$ when acting on the $m^{th}$ particle, and the $U(1)$ charge operator $Q_{U(1)}$ with eigenvalues $q_m$. Because  $Q^I_\alpha$ and $\overline{Q}^I_{\dot{\alpha}}$ commute with all these ingredients, it follows that the supercharge commutes with the leading soft photon operator in its entirety
\begin{equation}
    \Big[\mathcal{S}^{(-1)}_{\text{photon},+}(\widehat{p}_s),Q^I_{\alpha}\Big] = 0 \hspace{50pt} \Big[\mathcal{S}^{(-1)}_{\text{photon},+}(\widehat{p}_s),\overline{Q}^I_{\dot \alpha}\Big] = 0.
\end{equation}

The story gets more interesting at the subleading order, because now the angular momentum generator $J^{\mu \nu}$ -- which $Q^I_{\alpha}$ and $\overline{Q}^I_{\dot{\alpha}}$ do not commute with -- appears in the soft operator. We compute
\begin{equation}
    \hspace{-4pt}\Big[\mathcal{S}^{(0)}_{\text{photon},+}(\widehat{p}_s),Q^I_\alpha\Big] = - i e \sumop_{m=1}^n q_m \frac{\widehat{p}_{s} \cdot \big[J_m,Q^I_\alpha\big] \cdot \varepsilon_+(\widehat{p_s})}{\widehat{p}_s \cdot p_m} = i e\sumop_{m=1}^n q_m \frac{\widehat{p}_{s,\mu} \hspace{2pt} (\sigma^{\mu \nu})_{\alpha}^{~ \gamma} \hspace{2pt} \varepsilon_{+,\nu}(\widehat{p_s})}{\widehat{p}_s \cdot p_m} Q^I_{m, \gamma}.
\end{equation}
In this way, we see that the commutator is generally non-vanishing; it is responsible for the photino's leading soft theorem! One can act once more with the supercharge $Q^J_{\beta}$ on this new soft operator to determine the soft theorem for the scalar field in the supermultiplet. This time, the non trivial commutator is between $Q^I_{m,\gamma}$ and $Q^J_{\beta}$, and result is proportional to the central charge $Z^{IJ}_m$. 

Of course, this calculation is not the full story. One still needs to evaluate the commutator with the theory-dependent particle changing operators, $\mathcal{F}_{k,m}.$ However, this is a straightforward exercise (see footnote \ref{fnL particle changing}), and we will leave these expressions presented in a formal way defining $\mathcal{F}_{k,m}^{I_1 \cdots I_n}$ in analogy with $\mathcal{S}^{I_1 \cdots I_n}$ in Equation \eqref{eqn: soft theorem supermultiplet}
\begin{equation}
    \mathcal{F}^{I_1 \cdots I_n}_{k,m} =  \bigg(\prod_{m=1}^n \frac{1}{\sqrt{2}} \frac{[\xi_m|^{\alpha_m}}{ [ \xi_m\hspace{2pt} \widehat{p}_s]}\bigg) \hspace{3pt} \Big[\Big[\Big[\mathcal{F}_{k,m},Q^{I_n}_{\alpha_n}\Big]\hspace{2pt},\cdots\Big]\hspace{2pt},Q^{I_1}_{\alpha_1}\Big].
\end{equation}

In Appendix \ref{appendix: commutators}, we compute the full set of non-trivial commutators between supercharges and the universal parts of the soft photon and graviton theorems at leading few orders. The calculations involve a single line of algebra, and expressions are given for both positive and negative helicity sectors.

\subsection{List of Supersymmetric Soft Theorems in SQED and SUGRA}

We are now ready to completely determine the soft operators at leading orders for supersymmetric gauge theory and supergravity (including gauged supergravity). We define creation operators for superpartners of the photon and graviton using the following conventions:
\begin{table}[h]
\centering
\begin{tabular}{||l|c||l|c||} 
\hhline{|t:====:t|}
\multicolumn{4}{||c||}{\textbf{Gauge Multiplet}}                                          \\ 
\hhline{|:==:t:==:|}
\multicolumn{1}{||c|}{\textbf{Particle}} & \textbf{Helicity} & \multicolumn{1}{c|}{\textbf{Particle}} & \textbf{Helicity}  \\ 
\hhline{|:==::==:|}
   $\hspace{21pt}a_{\text{photon},+}(p) = a(p) \hspace{21pt}$                             & $+1$       &    $\hspace{21pt}a_{\text{photon},-}(p) = \overline{a}(p) \hspace{21pt}$                           & $-1$        \\ 
\hline
   $\hspace{17pt} a_{\text{photino},+}^I(p) = a^I(p) \hspace{17pt}$                             & $+\tfrac{1}{2}$        &   $\hspace{17pt}a_{\text{photino},-}^I(p) = \overline{a}^I(p) \hspace{17pt}$                            & $-\tfrac{1}{2}$         \\ 
\hline
   $\hspace{24pt}a_{\text{scalar},+}^{IJ}(p) = a^{IJ}(p)$                             & 0        &    $\hspace{24pt}a_{\text{scalar},-}^{IJ}(p) = \overline{a}^{IJ}(p)$                             & 0         \\
\hhline{|b:==:b:==:b|}
\end{tabular}
\\[-4pt]
\begin{tabular}{||l|c||l|c||} 
\hhline{|t:====:t|}
\multicolumn{4}{||c||}{\textbf{Gravity Multiplet}}                                          \\ 
\hhline{|:==:t:==:|}
\multicolumn{1}{||c|}{\textbf{Particle}} & \textbf{Helicity} & \multicolumn{1}{c|}{\textbf{Particle}} & \textbf{Helicity}  \\ 
\hhline{|:==::==:|}
  $\hspace{13pt}a_{\text{graviton},+}(p) = a(p)$                              & $+2$       &    $\hspace{13pt}a_{\text{graviton},-}(p) = \overline{a}(p)$                           & $-2$        \\ 
\hline
   \hspace{11pt}$a_{\text{gravitino},+}^I(p) = a^I(p)$                             & $+\tfrac{3}{2}$        &   \hspace{11pt}$a_{\text{gravitino},-}^I(p) = \overline{a}^I(p)$                            & $-\tfrac{3}{2}$         \\ 
\hline
   $a_{\text{graviphoton},+}^{IJ}(p) = a^{IJ}(p)$                             & $+1$       &    $a_{\text{graviphoton},-}^{IJ}(p) = \overline{a}^{IJ}(p)$                             & $-1$         \\
\hline
   $\hspace{16pt}a_{\text{dilatino},+}^{IJK}(p) = a^{IJK}(p)$                             & $+\tfrac{1}{2}$        &    $\hspace{16pt}a_{\text{dilatino},-}^{IJK}(p) = \overline{a}^{IJK}(p)$                             & $-\tfrac{1}{2}$         \\
\hline
   $\hspace{23pt}a_{\text{scalar},+}^{IJKL}(p) = a^{IJKL}(p)$                             & 0        &    $\hspace{23pt}a_{\text{scalar},-}^{IJKL}(p) = \overline{a}^{IJKL}(p)$                             & 0         \\
\hhline{|b:==:b:==:b|}
\end{tabular}
\caption{Our conventions for how states in supersymmetric QED and supergravity are related by successive action with the supercharges. \textbf{Caution:} Sometimes, factors of $i$ are included in these normalization conventions. One must keep track of them.}
\label{tab: particles in supermultiplet}
\end{table}

\newpage

One may read off the soft theorems associated to these particles by comparing the commutators in Appendix \ref{appendix: commutators} with Equation \eqref{eqn: soft theorem supermultiplet}. The results are given in Table \ref{fig: gravity soft theorems}. While we have only provided expressions for positive helicity particles, the negative helicity soft theorems are related by CPT conjugation:
\begin{equation}
    \big[\widehat{p}_s \hspace{2pt} Q^I_m\big] ~ \leftrightarrow ~ \big\langle \overline{Q}^I_m \hspace{2pt} \widehat{p}_s\big\rangle  ~ ~ ~ , ~ ~ ~ Z^{IJ} ~ \leftrightarrow ~ \overline{Z}^{IJ} ~ ~ ~ , ~ ~ ~ \varepsilon_+(\widehat{p}_s) ~ \leftrightarrow ~ \varepsilon_-(\widehat{p}_s) ~ ~ ~ , ~ ~ ~ g_k \mathcal{F}_k^{I_1 \cdots I_n} ~ \leftrightarrow ~ \overline{g}_k \overline{\mathcal{F}}_k^{\hspace{1pt}I_1 \cdots I_n}
\end{equation}
Note also that the soft theorems are manifestly antisymmetric in the $I_1,...,I_n$ indices as is necessary to be consistent with the total antisymmetry of the indices on $a^{I_1 \cdots I_n}(p_s)$. 

In Appendix \ref{appendix: comparison with literature}, we show that these results agree with the pioneering calculation for the soft photino theorem in $\mathcal{N} = 1$ supersymmetric QED with a charged, massless chiral multiplet \cite{Dumitrescu:2015fej}.

\begin{table}
\centering
\begin{tabular}{||l||l||c||} 
\hhline{|t:===:t|}
\multicolumn{3}{||c||}{\textbf{Soft Theorems in Supersymmetric Gauge Theories and Supergravity}}                                          \\ 
\hhline{|:=::=::=:|}
\multicolumn{1}{||c||}{\textbf{Operator}} & \multicolumn{1}{c||}{\textbf{Expression}} & \textbf{Order}             \\ 
\hhline{|:=::=::=:|}
\vcell{$\mathcal{S}_{\text{photon},+}$}                         & \vcell{$\mathlarger{e \hspace{2pt} \omega_s^{-1} \sumop_{m=1}^n q_m \hspace{2pt} \frac{\varepsilon_+ \cdot p_m}{\widehat{p}_s \cdot p_m}}$}                         & \vcell{$\omega_s^{-1}$}          \\[-29pt]
\printcellbottom                 & \printcellbottom                 & \printcellbottom  \\
\vcell{}                         & \vcell{\hspace{30pt}$\mathlarger{- i e \sumop_{m=1}^n q_m \hspace{2pt} \frac{\widehat{p}_{s} \cdot J_m \cdot \varepsilon_+}{\widehat{p}_s \cdot p_m} + g_k \sumop_{m=1}^n \hspace{2pt} \frac{[\hspace{2pt} \widehat{p}_s \hspace{2pt} p_m]}{\langle\hspace{2pt} \widehat{p}_s \hspace{2pt} p_m\rangle} \mathcal{F}_{k,m}}$}                         & \vcell{$\omega_s^{0}$}          \\[-33pt]
\printcellbottom                 & \printcellbottom                 & \printcellbottom  \\
\hline
\vcell{$\mathcal{S}^{I}_{\text{photino},+}$}                         & \vcell{$\mathlarger{-\frac{e}{2} \hspace{2pt} \omega_s^{-1/2} \sumop_{m=1}^n q_m \hspace{2pt} \frac{\big[ \widehat{p}_s \hspace{2pt} Q^I_{m}\big]}{\widehat{p}_s \cdot p_m} + g_k \hspace{2pt} \omega_s^{-1/2} \sumop_{m=1}^n \hspace{2pt} \frac{[\hspace{2pt} \widehat{p}_s \hspace{2pt} p_m]}{\langle\hspace{2pt} \widehat{p}_s \hspace{2pt} p_m\rangle} \mathcal{F}_{k,m}^I}$}                         & \vcell{$\omega_s^{-1/2}$}          \\[-31pt]
\printcellbottom                 & \printcellbottom                 & \printcellbottom  \\ 
\hline
\vcell{$\mathcal{S}^{IJ}_{\text{scalar},+}$}                         & \vcell{$\mathlarger{\frac{e}{2\sqrt{2}} \hspace{2pt} \omega_s^{-1} \sumop_{m=1}^n q_m \hspace{2pt}\frac{Z^{IJ}_m}{\widehat{p}_s \cdot p_m} + g_k \hspace{2pt} \omega_s^{-1}\sumop_{m=1}^n \hspace{2pt} \frac{[\hspace{2pt} \widehat{p}_s \hspace{2pt} p_m]}{\langle\hspace{2pt} \widehat{p}_s \hspace{2pt} p_m\rangle}\mathcal{F}_{k,m}^{IJ}}$}                         & \vcell{$\omega_s^{-1}$}          \\[-29pt]
\printcellbottom                 & \printcellbottom                 & \printcellbottom  \\
\hhline{|:=::=::=:|}
\vcell{$\mathcal{S}_{\text{graviton},+}$}                         & \vcell{$\mathlarger{\frac{\kappa}{2} \hspace{2pt} \omega_s^{-1} \sumop_{m=1}^n \frac{\big(p_m \cdot \varepsilon_+(\widehat{p}_s)\big)^2}{\widehat{p}_s \cdot p_m}}$}                         & \vcell{$\omega_s^{-1}$}          \\[-29pt]
\printcellbottom                 & \printcellbottom                 & \printcellbottom  \\
\vcell{}                         & \vcell{\hspace{30pt}$\mathlarger{- \frac{i\kappa}{2} \sumop_{m=1}^n \frac{\big(p_m \cdot \varepsilon_+(\widehat{p}_s)\big) \big(\widehat{p}_s \cdot J_m \cdot \varepsilon_+(\widehat{p}_s)\big)}{\widehat{p}_s \cdot p_m}}$}                         & \vcell{$\omega_s^0$}          \\[-31pt]
\printcellbottom                 & \printcellbottom                 & \printcellbottom  \\
\vcell{}                         & \vcell{\hspace{30pt}$\mathlarger{- \frac{\kappa}{4} \hspace{2pt} \omega_s \sumop_{m=1}^n \frac{\big(\widehat{p}_s \cdot J_m \cdot \varepsilon_+(\widehat{p}_s)\big)^2}{\widehat{p}_s \cdot p_m} + g_k \hspace{2pt} \omega_s  \sumop_{m=1}^n \frac{[\hspace{2pt} \widehat{p}_s \hspace{2pt} p_m]^3}{\langle\hspace{2pt} \widehat{p}_s \hspace{2pt} p_m\rangle} \mathcal{F}_{k,m}}$}                         & \vcell{$\omega_s$}          \\[-31pt]
\printcellbottom                 & \printcellbottom                 & \printcellbottom  \\ 
\hline
\vcell{$\mathcal{S}^{I}_{\text{gravitino},+}$}                         & \vcell{$\mathlarger{-\frac{\kappa}{4} \hspace{2pt} \omega_s^{-1/2} \sumop_{m=1}^n \frac{p_m \cdot \varepsilon_+}{\widehat{p}_s \cdot p_m}  \hspace{2pt} \big[\widehat{p}_s \hspace{2pt} Q^I_{m}\big]}$}                         & \vcell{$\omega_s^{-1/2}$}          \\[-29pt]
\printcellbottom                 & \printcellbottom                 & \printcellbottom  \\
\vcell{}                         & \vcell{\hspace{30pt}$\mathlarger{+\frac{i\kappa}{8}\hspace{2pt} \omega_s^{1/2} \sumop_{m=1}^n \frac{\big[\widehat{p}_s \hspace{2pt} Q^I_m\big] \hspace{2pt} (\widehat{p}_s \cdot J_m \cdot \varepsilon_+) +(\widehat{p}_s \cdot J_m \cdot \varepsilon_+) \big[\widehat{p}_s \hspace{2pt} Q^I_m\big]}{\widehat{p}_s \cdot p_m}}$}                         & \vcell{$\omega_s^{1/2}$}          \\[-31pt]
\printcellbottom                 & \printcellbottom                 & \printcellbottom  \\
\vcell{}                         & \vcell{\hspace{30pt}$\mathlarger{+ g_k \hspace{2pt} \omega_s^{1/2} \sumop_{m=1}^n \frac{[\hspace{2pt} \widehat{p}_s \hspace{2pt} p_m]^3}{\langle\hspace{2pt} \widehat{p}_s \hspace{2pt} p_m\rangle} \mathcal{F}_{k,m}^I}$}                         & \vcell{$\omega_s^{1/2}$}          \\[-31pt]
\printcellbottom                 & \printcellbottom                 & \printcellbottom  \\ 
\hline
\vcell{$\mathcal{S}^{IJ}_{\text{graviphoton},+}$}                         & \vcell{$\mathlarger{\frac{\kappa}{4\sqrt{2}} \hspace{2pt} \omega_s^{-1} \sumop_{m=1}^n \frac{p_m \cdot \varepsilon_+}{\widehat{p}_s \cdot p_m} \hspace{2pt} Z_m^{IJ}}$}                         & \vcell{$\omega_s^{-1}$}          \\[-29pt]
\printcellbottom                 & \printcellbottom                 & \printcellbottom  \\
\vcell{}                         & \vcell{\hspace{30pt}$\mathlarger{+\frac{\kappa}{8} \sumop_{m=1}^n \frac{\big[\widehat{p}_s \hspace{2pt} Q^{[I}_m\big]\big[\widehat{p}_s \hspace{2pt} Q^{J]}_m\big] - \sqrt{2} i \hspace{2pt}(\widehat{p}_s \cdot J_m \cdot \varepsilon_+) \hspace{2pt} Z_m^{IJ}}{\widehat{p}_s \cdot p_m}}$}                         & \vcell{$\omega_s^{0}$}          \\[-33pt]
\printcellbottom                 & \printcellbottom                 & \printcellbottom  \\
\vcell{}                         &\vcell{\hspace{30pt}$\mathlarger{+ g_k \sumop_{m=1}^n \frac{[\hspace{2pt} \widehat{p}_s \hspace{2pt} p_m]^3}{\langle\hspace{2pt} \widehat{p}_s \hspace{2pt} p_m\rangle} \mathcal{F}_{k,m}^{IJ}}$}                         & \vcell{$\omega_s^{0}$}          \\[-33pt]
\printcellbottom                 & \printcellbottom                 & \printcellbottom  \\ 
\hline
\vcell{$\mathcal{S}^{IJK}_{\text{dilatino},+}$}                         & \vcell{$\mathlarger{-\frac{3\kappa}{8 \sqrt{2}} \hspace{2pt} \omega_s^{-1/2} \sumop_{m=1}^n \frac{\big[ \widehat{p}_s \hspace{2pt} Q^{[I}_m\big] \hspace{2pt} Z_m^{JK]}}{\widehat{p}_s \cdot p_m} + g_k\hspace{2pt} \omega_s^{-1/2} \sumop_{m=1}^n \frac{[\hspace{2pt} \widehat{p}_s \hspace{2pt} p_m]^3}{\langle\hspace{2pt} \widehat{p}_s \hspace{2pt} p_m\rangle} \mathcal{F}_{k,m}^{IJK}}$}                         & \vcell{$\omega_s^{-1/2}$}          \\[-31pt]
\printcellbottom                 & \printcellbottom                 & \printcellbottom  \\ 
\hline
\vcell{$\mathcal{S}^{IJKL}_{\text{scalar},+}$}                         & \vcell{$\mathlarger{\frac{3 \kappa}{16} \hspace{2pt} \omega_s^{-1} \sumop_{m=1}^n \frac{Z_m^{[IJ} Z_m^{KL]}}{\widehat{p}_s \cdot p_m}  +g_k \hspace{2pt} \omega_s^{-1} \sumop_{m=1}^n \frac{[\hspace{2pt} \widehat{p}_s \hspace{2pt} p_m]^3}{\langle\hspace{2pt} \widehat{p}_s \hspace{2pt} p_m\rangle} \mathcal{F}_{k,m}^{IJKL}}$}                         & \vcell{$\omega_s^{-1}$}          \\[-29pt]
\printcellbottom                 & \printcellbottom                 & \printcellbottom  \\
\hhline{|b:=:b:=:b:=:b|}
\end{tabular}
\caption{Soft theorems for particles listed in Table \ref{tab: particles in supermultiplet}. These comprise the leading order soft theorems for all particles in the vector multiplet of supersymmetric gauge theories and the gravity multiplet of supergravity theories. The expressions are valid for gauged supergravity as well.}
\label{fig: gravity soft theorems}
\end{table}

\subsection{Observations on Supersymmetric Soft Theorems in Gauge Theory and Gravity}

We conclude this section with a few remarks about our findings. First, we note that \textit{the leading soft photino theorem is not universal}, which contradicts the claims of \cite{Dumitrescu:2015fej}. Indeed, there are theory dependent operators $\mathcal{F}_{k}$ which may appear in the subleading soft photon theorem and contribute to the leading soft photino theorem. In Appendix \ref{appendix: comparison with literature}, we show that our results are consistent with \cite{Dumitrescu:2015fej}, however, because its authors consider a model where such non-universal corrections are absent, subverting this potential pitfall. 

The \textit{leading soft gravitino theorem, however, is universal}! This is important for the celestial holography program because the leading soft gravitino theorem is responsible for generating a $(\tfrac{3}{2},0)$ current on the celestial sphere which enhances the conformal symmetry of the celestial dual to a super BMS symmetry \cite{Fotopoulos:2020bqj, Awada:1985by, Henneaux:2020ekh,Fuentealba:2021xhn,Banerjee:2022abf,Boulanger:2023gpw, Banerjee:2022lnz}. Because the leading soft gravitino theorem plays such a pivotal role in celestial holography, it is heartening to firmly conclude that it is not theory dependent.

Another interesting observation is related to the order of the pole in the soft expansion. From Equation \eqref{eqn: soft theorem supermultiplet}, one observes that every time one acts on a soft operator with a supercharge to get a new soft operator, the pole of the new soft operator is more singular by a factor of $\omega_s^{-1/2}.$ On the other hand, acting with a supercharge on the leading soft theorem gives a trivial commutator, as does acting with more than two supercharges on the subleading soft theorem. Thus, as you act with more supercharges, one must go deeper into the soft expansion to obtain a non-trivial result -- this generates positive powers of $\omega_s.$ These powers of $\omega_s^{-1/2}$ and $\omega_s$ balance such that the leading soft theorem for bosons occurs at $\mathcal{O}(\omega_s^{-1})$ while the leading soft theorem for fermions occurs at $\mathcal{O}(\omega_s^{-1/2})$, at least in the case of gauge and gravitational theories.

Finally, we remark that a photon may appear in a supermultiplet and not be the highest-weight state. This is the case in supergravity for example, which features a $U(1)$ gauge boson -- the graviphoton -- obtained by acting on the graviton with two supercharges. Thus, the soft graviphoton theorem needs to match the soft photon theorem (whose leading term is universal). Comparing the leading soft photon and graviphoton theorems necessitates the relationship $4 \sqrt{2} \hspace{2pt} e \hspace{1pt} q_m = \kappa \hspace{1pt} Z^{IJ}_m$; fortunately, the central charge of a particle is related to its $U(1)$ charge in precisely this way, so this is consistent \cite{Freedman:2012zz}.\footnote{Note the central charge of a particle with both electric and magnetic charge is $Z^{IJ} = 4 \sqrt{2} \hspace{1pt}e \hspace{1pt} \kappa^{-1} (q + 2\pi i g/e^2).$ Fortunately, the soft photon theorem has also been studied in the presence of external magnetically charged particles \cite{Strominger:2015bla}. The deformation to the soft photon theorem again precisely matches the above!}

\section{Discussion}
\label{sec: discussion}

In this work, we have shown how the supersymmetric Ward identity can be leveraged to compute soft theorems for a particle's entire supermultiplet -- the expressions are compact, and the derivations involve only a single line of algebra. As a proof of concept, we have illustrated the utility of this formalism by deriving soft theorems for the entire gauge and gravity multiplets in 4d supersymmetric theories. Moreover, this framework is completely generalizable, reducing the previously arduous problem of computing soft theorems with Feynman diagrams to evaluating commutators with the supersymmetry algebra. We conclude by highlighting some directions of inquiry which this formalism may illuminate.

\textbf{Arbitrary Symmetry Algebras:} The story that we have presented so far concerns the supersymmetry algebra and its representation theory. A quantum theory may have a richer symmetry algebra, such as superconformal symmetry with new generators $S_{I \alpha}$ and $K_\mu$. Now, one can move between annihilation operators and soft theorems in a superconformal multiplet in a similar manner. For example, if $a_{I\alpha}(p) = \big[a(p),S_{I\alpha}\big]$, then one can immediately deduce that their soft theorems satisfy $\mathcal{S}_{I\alpha}(p) = \big[\mathcal{S}(p),S_{I \alpha}\big]$. This is an extremely powerful constraint. 

More broadly, if $\frak{g}$ is \textit{any symmetry algebra} of some theory (which may include both spacetime and internal symmetries), then the particles organize themselves into a representation of $\frak{g}$, so the corresponding soft theorems must as well. It would be intriguing to use this toolkit to study other interesting theories with physically relevant $\frak{g}.$

\textbf{Asymptotic Symmetries:} It is known that soft theorems imply the existence of infinitely-many conserved charges, $\mathcal{Q}[f]$, which are parameterized by an arbitrary function $f(z,\bar z)$ living on the celestial sphere. The charges are typically decomposed into the sum of hard and soft pieces
\begin{equation}
    \mathcal{Q}[f] = \mathcal{Q}_H[f] + \mathcal{Q}_S[f]. 
\end{equation}
The hard part of the conserved charge, $\mathcal{Q}_H[f]$, is responsible for measuring various fluxes of external hard particles in a scattering amplitude (for example, the charge or energy of a hard particle weighted by the location that particle exits the celestial sphere). Explicit expressions for these conserved charges have been written down in terms of bulk field variables like the electromagnetic field strength or the metric tensor. Conversely, the soft part of the conserved charge, $\mathcal{Q}_S[f]$, is responsible for adding some distribution of soft particles -- generally, $\mathcal{Q}_S[f]$ looks like the creation/annihilation operator for some soft particle at a particular order in the soft expansion integrated against a kernel parameterized by the function $f.$

It is clear that if the charges $\mathcal{Q}[f]$ are conserved (i.e. commute with the S-matrix), then the charge $\mathcal{Q}^I_{\alpha}[f] = \big[\mathcal{Q}[f],Q^I_{\alpha}\big]$ will be similarly conserved. This charge also factorizes into the sum of hard and soft parts where $\mathcal{Q}^I_{\alpha,H}[f]$ can be directly computed from the explicit expression for $\mathcal{Q}_H[f]$ and $\mathcal{Q}^I_{\alpha,S}[f]$ will be an integral of the creation/annihilation operator for the superpartner of the original soft particle. 

One can apply this procedure successively to define the conserved charges $\mathcal{Q}^{I_1 \cdots I_n}_{\alpha_1 \cdots \alpha_n}[f]$ which must also live in a representation of the supersymmetry algebra. It would be interesting to study these supermultiplets of conserved charges in greater detail and to examine in what sense their corresponding asymptotic symmetries also organize themselves under the action of supersymmetry.

\textbf{Celestial Holography:} One can also use this formalism to study general aspects of celestial holography with spacetime supersymmetry. There have been some results in this direction, though, again, the authors tend to work with a particular Lagrangian from the start. One can hope to understand how conformally soft theorems, OPEs, and holographic symmetry algebras are organized in supersymmetric theories more broadly. This is achieved in the upcoming work \cite{Tropper:2024evi}.

\textbf{Soft Theorems and Geometry:} It has been  shown that soft theorems for scalars can be related to geometric data appearing in the Lagrangian of a theory \cite{Cheung:2021yog, Derda:2024jvo}. For example, in a non-linear sigma model with metric $g_{IJ}(\Phi)$, there is a geometric soft theorem:
\begin{equation}
    \langle 0| a_{I_1}(p_1) \cdots a_{I_n}(p_n) \hspace{1pt} a_{I_s}(p_s) |0\rangle_v \xrightarrow{p_s \hspace{2pt} \rightarrow \hspace{2pt} 0} \nabla_{I_s} \langle 0| a_{I_1}(p_1) \cdots a_{I_n}(p_n) |0\rangle_v,
\end{equation}
where the amplitudes are computed at some vev $\Phi^I = v^I$ for the scalars and $\nabla_{I_s}$ is the covariant derivative with respect to the non-linear sigma model metric acting on the scattering amplitude as a function of the moduli, $v^I$. Thus, soft theorems encode the geometry of the moduli space of vacua; this has applications to the celestial holography program studied in \cite{Kapec:2022axw}. Supersymmetric non-linear sigma models have a K\"ahler metric, and it would be interesting to determine what quantities in the K\"ahler geometry the fermionic soft theorems are measuring.

Another interesting class of examples is afforded by studying Seiberg-Witten theories. These are $\mathcal{N} = 2$ gauge theories which have rich moduli spaces, the geometry of which is captured by various soft theorems. In the upcoming work \cite{WIP:Seiberg-Witten}, we study how this geometry is manifested in the celestial dual to such theories and the constraining relationships that supersymmetry imposes.

\subsection*{Acknowledgements}

I would like to thank Erin Crawley, Alfredo Guevara, Matt Heydeman, Marcus Spradlin, Andrew Strominger, Tomasz Taylor, Chiara Toldo, and Anastasia Volovich for many useful discussions and helpful comments on the draft. This work is supported by NSF GRFP grant DGE1745303.

\appendix
\section{Supersymmetric Soft Theorems in On-Shell Superspace}
\label{appendix: on shell superspace}

The relation among soft theorems in a supermultiplet takes an especially elegant form, when written in the language of \textit{on-shell superspace} (for introductions to on-shell superspace, see \cite{Nair:1988bq, Elvang:2011fx,Elvang:2013cua,Tropper:2024evi}). This was originally noticed in \cite{Liu:2014vva} for the case of $\mathcal{N} = 4$ super Yang-Mills and $\mathcal{N} = 8$ supergravity. In this section, we broaden their results to a more general setting.

We begin by introducing $\mathcal{N}$ real Grassmann variables, $\eta_1,...,\eta_{\mathcal{N}}$ which allow us to package all particles in the $a(p)$ supermultiplet into a single on-shell superfield $\mathbb{A}(p|\eta)$
\begin{equation}
    \mathbb{A}(p|\eta) = \sumop_{n=0}^\mathcal{N} \frac{1}{n!} \hspace{1pt} \eta_{I_1}\cdots \eta_{I_n} \hspace{1pt} a^{I_1 \cdots I_n}(p)\hspace{2pt},
    \label{eqn: on shell supermultiplet}
\end{equation}
obeying
\begin{equation}
    \Big[\mathbb{A}(p|\eta),Q^I_\alpha\Big] = \sqrt{2}\hspace{2pt} |p]_{\alpha} \frac{\partial}{\partial \eta_I} \mathbb{A}(p|\eta) \hspace{40pt} \Big[\mathbb{A}(p|\eta),\overline{Q}^I_{\dot\alpha}\Big] = \sqrt{2}\hspace{2pt} \langle p|_{\dot \alpha} \hspace{2pt} \eta_I \hspace{1pt} \mathbb{A}(p|\eta)\hspace{2pt}. 
    \label{eqn: supercharge commutator with superoperator}
\end{equation}
The power of the on-shell superspace formalism is that it allows one to package a collection of scattering amplitudes into a single \textit{superamplitude}. Indeed, all $n$-point functions involving particles in the supermultiplets generated by $a_{j_1}(p),...,a_{j_n}(p)$ are encapsulated in the $\langle 0| \mathbb{A}_{j_1}(p_1|\eta_1) \cdots \mathbb{A}_{j_n}(p_n|\eta_n)|0\rangle$ superamplitude. Individual component amplitudes can be extracted via appropriate Grassmann derivatives. 

We can now generalize Equation \eqref{eqn: soft theorem definition} by formally writing the soft theorem in superspace
\begin{equation}
    \langle 0| \mathbb{A}_{j_1}(p_1|\eta_1) \cdots \mathbb{A}_{j_n}(p_n|\eta_n) \mathbb{A}(p_s|\eta_s) |0\rangle \xrightarrow{p_s \rightarrow 0} \langle 0| \mathbb{A}_{j_1}(p_1|\eta_1) \cdots \mathbb{A}_{j_n}(p_n|\eta_n) |0\rangle \bullet \mathbb{S}(p_s|\eta_s)\hspace{2pt},
\end{equation}
where we have defined the soft operator in superspace to match Equation \eqref{eqn: on shell supermultiplet}
\begin{equation}
    \mathbb{S}(p|\eta) =  \sumop_{n=0}^\mathcal{N} \frac{1}{n!} \hspace{1pt} \eta_{I_1}\cdots \eta_{I_n} \hspace{1pt} \mathcal{S}^{I_1 \cdots I_n}(p)\hspace{2pt}.
\end{equation}
Equation \eqref{eqn: soft theorem SUSY representation} implies that such operators obey
\begin{equation}
    \Big[\mathbb{S}(p|\eta),Q^I_\alpha\Big] = \sqrt{2}\hspace{2pt} |p]_{\alpha} \frac{\partial}{\partial \eta_I} \mathbb{S}(p|\eta) \hspace{40pt} \Big[\mathbb{S}(p|\eta),\overline{Q}^I_{\dot\alpha}\Big] = \sqrt{2}\hspace{2pt} \langle p|_{\dot \alpha} \hspace{2pt} \eta_I \hspace{1pt} \mathbb{S}(p|\eta)\hspace{2pt}. 
\end{equation}
Again, the structure of this equation matches Equation \eqref{eqn: supercharge commutator with superoperator}. This is an equivalent way of stating our earlier result that soft theorems form a representation of the supersymmetry algebra. The language of on-shell superspace allows these expressions to take a particularly beautiful and compact form because all equations in \eqref{eqn: soft theorem SUSY representation} are now combined into a single expression written in terms of a function on superspace.

Of course, similar expressions can be obtained for soft theorems associated to particles in the CPT conjugate supermultiplet.

\section{Supersymmetric Soft Theorems in Arbitrary Dimensions}
\label{appendix: arbitrary dimensions}

In arbitrary dimensions, the story is extremely similar. One can consider a theory in $d$ dimensions with $\mathcal{N}$ supercharges $Q^1_\alpha,\cdots, Q^\mathcal{N}_{\alpha}.$ If $a(p)$ is the annihilation operator for some state in the theory, it is always true that the remaining states in the supermultiplet can be constructed by successively acting on $a(p)$ with those supercharges. This means that the supermultiplet is spanned by single particle states whose annihilation operators take the form\footnote{Note that annihilation operators defined in such a way will neither be linearly independent, nor will they satisfy the canonical commutation relations $[a^{I_1 \cdots I_n}_{\alpha_1 \cdots \alpha_n}(p),a^{I_1 \cdots I_n \dagger}_{\alpha_1 \cdots \alpha_n}(p')] = 2E_p \delta^{(d-1)}(\vec{p} - \vec{p}')$. By using the SUSY algebra in the dimension of interest, one may infer such features.}
\begin{equation}
    a^{I_1 \cdots I_n}_{\alpha_1 \cdots \alpha_n}(p) = \Big[\Big[\Big[a(p),Q^{I_n}_{\alpha_n}\Big]\hspace{2pt},\cdots\Big]\hspace{2pt}, Q^{I_1}_{\alpha_1}\Big].
\end{equation}
To deduce the soft theorem for $a^{I_1 \cdots I_n}_{\alpha_1 \cdots \alpha_n}(p)$ in terms of $a(p)$, one simply uses the SUSY Ward identity, repeating the derivation in Section \ref{sec: SUSY soft theorems}. Concretely, if the particle $a(p)$ is associated to the soft operator $\mathcal{S}(p)$, then the particle $a^{I_1 \cdots I_n}_{\alpha_1 \cdots \alpha_n}(p)$ is associated to the soft operator $\mathcal{S}^{I_1 \cdots I_n}_{\alpha_1 \cdots \alpha_n}(p)$, where
\begin{equation}
    \mathcal{S}^{I_1 \cdots I_n}_{\alpha_1 \cdots \alpha_n}(p) = \Big[\Big[\Big[\mathcal{S}(p),Q^{I_n}_{\alpha_n}\Big]\hspace{2pt},\cdots\Big]\hspace{2pt},Q^{I_1}_{\alpha_1}\Big].
\end{equation}

\section{Explicit Expressions for the Commutators}
\label{appendix: commutators}

In this appendix, we compute the commutators between the supercharges and the \textit{universal parts} (denoted with a tilde) of the soft photon and graviton theorems at leading orders in the soft expansion. We have already shown that the leading soft positive helicity photon theorem has a trivial commutator with the supercharges. The same conclusion holds for negative helicity photons as well as gravitons of both helicities
\begin{equation}
    \begin{split}
        \Big[\widetilde{S}^{(-1)}_{\text{photon},\pm}(\widehat{p}_s),Q^I_{\alpha}\Big] &= 0 \hspace{74pt} \Big[\widetilde{S}^{(-1)}_{\text{photon},\pm}(\widehat{p}_s),\overline{Q}^I_{\dot{\alpha}}\Big] = 0\\
        \Big[\widetilde{S}^{(-1)}_{\text{graviton},\pm}(\widehat{p}_s),Q^I_{\alpha}\Big] &= 0 \hspace{70pt} \Big[\widetilde{S}^{(-1)}_{\text{graviton},\pm}(\widehat{p}_s),\overline{Q}^I_{\dot{\alpha}}\Big] = 0.
    \end{split}
\end{equation}
To evaluate the commutators at subleading order, we need the following intermediate expressions which are determined directly from the supersymmetry algebra:
\begin{equation}
    \begin{split}
        \Big[J^{\mu \nu},Q^I_\alpha \Big] &= -(\sigma^{\mu \nu})_{\alpha}^{~ \gamma} \hspace{2pt} Q^I_{\gamma} \hspace{70pt} \Big[(\sigma^{\mu \nu})_{\alpha}^{~ \gamma}\hspace{2pt} Q^I_\gamma,Q^J_{\beta}\Big] = -\sigma^{\mu \nu}_{\alpha \beta} \hspace{2pt} Z^{IJ}\\
        \Big[J^{\mu \nu},\overline{Q}^I_{\dot{\alpha}}\Big] &= -(\overline{\sigma}^{\mu \nu})_{\dot{\alpha}}^{~ \dot{\gamma}}\hspace{2pt}\overline{Q}^I_{\dot \gamma} \hspace{70pt} \Big[(\overline{\sigma}^{\mu \nu})_{\dot \alpha}^{~ \dot \gamma} \hspace{2pt} \overline{Q}^I_{\dot \gamma},\overline{Q}^J_{\dot \beta}\Big] = - \overline{\sigma}^{\mu \nu}_{\dot \alpha \dot \beta} \hspace{2pt} \overline{Z}^{IJ}.
        \label{eqn: supercharge identities}
    \end{split}
\end{equation}
It will also be helpful to note the identities:
\begin{equation}
    \begin{split}
        \big(\widehat{p}_s \cdot \sigma\cdot \varepsilon_{+}(\widehat{p}_s)\big)\hspace{.1pt}_{\alpha\beta}&= \frac{i}{\sqrt{2}} |\widehat{p}_s]_{\alpha} \hspace{1pt} |\widehat{p}_s]_{\beta} \hspace{64pt} \big(\widehat{p}_s \cdot \sigma \cdot \varepsilon_{-}(\widehat{p}_s)\big)\hspace{.1pt}_{\alpha \beta} = 0\\
        \big(\widehat{p}_s \cdot \overline{\sigma}\cdot \varepsilon_{+}(\widehat{p}_s)\big)\hspace{.1pt}_{\dot \alpha \dot \beta} &= 0\hspace{122pt} \big(\widehat{p}_s \cdot \overline{\sigma}\cdot \varepsilon_{-}(\widehat{p}_s)\big)\hspace{.1pt}_{\dot \alpha \dot \beta} = \frac{i}{\sqrt{2}} \langle \widehat{p}_s|_{\dot{\alpha}}\langle \widehat{p}_s|_{\dot{\beta}}.
        \label{eqn: spinor-helicity identities}
    \end{split}
\end{equation}
These expressions imply that even at subleading order, certain supercharges commute with the soft theorems in their entirety
\begin{equation}
    \begin{split}
        \Big[\widetilde{\mathcal{S}}^{(k)}_{\text{photon},+}(\widehat{p}_s),\overline{Q}^I_{\dot{\alpha}}\Big] &= 0 \hspace{75pt} \Big[\widetilde{\mathcal{S}}^{(k)}_{\text{photon},-}(\widehat{p}_s),Q^I_{\alpha}\Big] = 0\\
        \Big[\widetilde{\mathcal{S}}^{(k)}_{\text{graviton},+}(\widehat{p}_s),\overline{Q}^I_{\dot \alpha} \Big] &= 0 \hspace{70pt} \Big[ \widetilde{\mathcal{S}}^{(k)}_{\text{graviton},-}(\widehat{p}_s),Q^I_{\alpha}\Big] = 0.
    \end{split}
\end{equation}
Note that there is a notion of a universal part of the soft photon and graviton theorems even when $k \geq 2$ for which these relations hold as well \cite{He:2014bga,Guevara:2019ypd}. This is a crucial consistency check on this formalism. Indeed, positive helicity photons and gravitons are also highest-weight states in the vector and gravity multiplets, respectively. This means they satisfy the highest-weight conditions $\big[a_{\text{photon},+},\overline{Q}^I_{\dot{\alpha}}\big] = \big[a_{\text{graviton},+},\overline{Q}^I_{\dot{\alpha}}\big] = 0$. Because there is no particle above them in the supermultiplet, the would-be soft theorem associated to the nonexistent larger-helicity state must vanish identically. This just means that the positive helicity soft photon and graviton theorems are highest-weight in their representation of the supersymmetry algebra (compare with Equation \eqref{eqn: soft theorem SUSY representation}). A similar story is true for the negative helicity photons and gravitons which satisfy the lowest-weight condition $\big[a_{\text{photon},-},Q^I_\alpha\big] = \big[a_{\text{graviton},-},Q^I_\alpha\big] = 0$ in the vector and gravity multiplets.

With these ingredients in hand, we are ready to compute the non-trivial commutators.

\subsubsection*{Commutators with $\widetilde{\mathcal{S}}^{(0)}_{\textmd{photon},\pm}(\widehat{p}_s)$:}

For the positive helicity photon:
\begin{align}
        \widetilde{\mathcal{S}}^{(0)}_{\text{photon},+}(\widehat{p}_s) &= - i e \sumop_{m=1}^n q_m \frac{(\widehat{p}_{s} \cdot J_m \cdot \varepsilon_+)}{\widehat{p}_s \cdot p_m} \nonumber \\
        \Big[\widetilde{\mathcal{S}}^{(0)}_{\text{photon},+}(\widehat{p}_s),Q^I_\alpha\Big] &=  +i e \sumop_{m=1}^n  q_m \frac{(\widehat{p}_{s} \cdot \sigma \cdot \varepsilon_+)_{\alpha}^{~ \gamma}}{\widehat{p}_s \cdot p_m} \hspace{2pt} Q^I_{m,\gamma} = -\frac{e}{\sqrt{2}} \bigg(\sumop_{m=1}^n \frac{q_m \big[\widehat{p}_s\hspace{2pt} Q^I_{m} \big]}{\widehat{p}_s \cdot p_m} \bigg) |\widehat{p}_s]_{\alpha} \\
        \Big[\Big[\widetilde{\mathcal{S}}^{(0)}_{\text{photon},+}(\widehat{p}_s),Q^I_{\alpha}\Big],Q^J_{\beta}\Big] &=  - i e \sumop_{m=1}^n  q_m \frac{(\widehat{p}_{s} \cdot \sigma \cdot \varepsilon_+)_{\alpha\beta}}{\widehat{p}_s \cdot p_m} \hspace{2pt} Z^{IJ}_m = +\frac{e}{\sqrt{2}} \bigg(\sumop_{m=1}^n \frac{q_m \hspace{2pt}Z^{IJ}_m}{\widehat{p}_s \cdot p_m} \bigg) |\widehat{p}_s]_{\alpha}|\widehat{p}_s]_{\beta}. \nonumber
\end{align}
Acting with a third supercharge will yield zero because $\big[Z^{IJ},Q^K_{\gamma}\big] = 0.$

For the negative helicity photon:
\begin{equation}
    \begin{split}
        \widetilde{\mathcal{S}}^{(0)}_{\text{photon},-}(\widehat{p}_s) &= - i e \sumop_{m=1}^n q_m \frac{(\widehat{p}_{s} \cdot J_m \cdot \varepsilon_-)}{\widehat{p}_s \cdot p_m} \hspace{2pt} \\
        \Big[\widetilde{\mathcal{S}}^{(0)}_{\text{photon},-}(\widehat{p}_s),\overline{Q}^I_{\dot{\alpha}}\Big] &=  + i e \sumop_{m=1}^n  q_m \frac{(\widehat{p}_{s} \cdot \overline{\sigma} \cdot \varepsilon_-)_{\dot \alpha}^{~ \dot \gamma}}{\widehat{p}_s \cdot p_m} \overline{Q}^I_{m,\dot\gamma} = -\frac{e}{\sqrt{2}} \bigg(\sumop_{m=1}^n \frac{q_m \big\langle \overline{Q}^I_m \widehat{p}_s\big\rangle}{\widehat{p}_s \cdot p_m} \hspace{2pt} \bigg) \langle \widehat{p}_s|_{\dot \alpha} \\
        \hspace{-1.5pt}\Big[\Big[\widetilde{\mathcal{S}}^{(0)}_{\text{photon},-}(\widehat{p}_s),\overline{Q}^I_{\dot{\alpha}}\Big],\overline{Q}^J_{\dot{\beta}}\Big] &=  - i e \sumop_{m=1}^n  q_m \frac{(\widehat{p}_{s} \cdot \overline{\sigma} \cdot \varepsilon_-)_{\dot\alpha\dot\beta} }{\widehat{p}_s \cdot p_m} \hspace{1pt} \overline{Z}^{IJ}_m = +\frac{e}{\sqrt{2}} \bigg(\sumop_{m=1}^n \frac{q_m \overline{Z}^{IJ}_m}{\widehat{p}_s \cdot p_m} \bigg) \langle \widehat{p}_s|_{\dot \alpha}\langle \widehat{p}_s|_{\dot \beta}. \\
    \end{split}
\end{equation}

We emphasize that these calculations are precisely one line long. One first uses Equation \eqref{eqn: supercharge identities} to evaluate the commutator, then one uses Equation \eqref{eqn: spinor-helicity identities}, if desired, to put the expression in the cleaner notation afforded by spinor-helicity variables.
 
\subsubsection*{Commutators with $\widetilde{\mathcal{S}}^{(0)}_{\textmd{graviton},\pm}(\widehat{p}_s)$:}

For the positive helicity graviton (this time going straight to spinor-helicity variables):
\begin{equation}
    \begin{split}
        \mathcal{S}^{(0)}_{\text{graviton},+}(\widehat{p}_s) &= -\frac{i\kappa}{2} \bigg(\sumop_{m=1}^n \frac{p_m \cdot \varepsilon_+}{\widehat{p}_s \cdot p_m} \hspace{2pt} (\widehat{p}_s \cdot J_m \cdot \varepsilon_+)\bigg) \\
        \Big[\widetilde{S}^{(0)}_{\text{graviton},+}(\widehat{p}_s),Q^I_\alpha \Big] &= -\frac{\kappa}{2\sqrt{2}} \bigg(\sumop_{m=1}^n \frac{p_m \cdot \varepsilon_+}{\widehat{p}_s \cdot p_m}  \hspace{2pt} \big[ \widehat{p}_s \hspace{2pt} Q^I_m\big] \bigg)|p_s]_{\alpha}\\
        \Big[\Big[\widetilde{S}^{(0)}_{\text{graviton},+}(\widehat{p}_s),Q^I_\alpha\Big],Q^J_\beta\Big] &= +\frac{\kappa}{2\sqrt{2}} \bigg(\sumop_{m=1}^n \frac{p_m \cdot \varepsilon_+}{\widehat{p}_s \cdot p_m} \hspace{2pt} Z_m^{IJ}\bigg) |\widehat{p}_s]_{\alpha} |\widehat{p}_s]_{\beta}.\\
    \end{split}
\end{equation}
For the negative helicity graviton:
\begin{equation}
    \begin{split}
        \mathcal{S}^{(0)}_{\text{graviton},-}(\widehat{p}_s) &= -\frac{i\kappa}{2} \bigg(\sumop_{m=1}^n \frac{p_m \cdot \varepsilon_-}{\widehat{p}_s \cdot p_m} \hspace{2pt} (\widehat{p}_s \cdot J_m \cdot \varepsilon_-)\bigg) \\
        \Big[\widetilde{\mathcal{S}}^{(0)}_{\text{graviton},-}(\widehat{p}_s),\overline{Q}^I_{\dot{\alpha}}\Big] &= -\frac{\kappa}{2\sqrt{2}} \bigg(\sumop_{m=1}^n \frac{p_m \cdot \varepsilon_-}{\widehat{p}_s \cdot p_m} \hspace{2pt} \big\langle \overline{Q}^I_m \hspace{2pt} \widehat{p}_s \big\rangle \bigg) \langle \widehat{p}_s |_{\dot \alpha}\\
        \Big[\Big[\widetilde{\mathcal{S}}^{(0)}_{\text{graviton},-}(\widehat{p}_s),\overline{Q}^I_{\dot\alpha}\Big],\overline{Q}^J_{\dot \beta}\Big] &= +\frac{\kappa}{2\sqrt{2}} \bigg( \sumop_{m=1}^n \frac{p_m \cdot \varepsilon_-}{\widehat{p}_s \cdot p_m} \hspace{2pt} \overline{Z}^{IJ}_{m}\bigg)\langle\widehat{p}_s|_{\dot \alpha} \langle \widehat{p}_s|_{\dot \beta}.\\
    \end{split}
\end{equation}

\subsubsection*{Commutators with $\widetilde{\mathcal{S}}^{(1)}_{\textmd{graviton},\pm}(\widehat{p}_s)$:}

For the positive helicity graviton, the following commutators: 
\begin{equation}
    \widetilde{\mathcal{S}}^{(1)}_{\text{graviton},+}(\widehat{p}_s) \hspace{8pt} , \hspace{8pt} \Big[ \widetilde{\mathcal{S}}^{(1)}_{\text{graviton},+}(\widehat{p}_s),Q^I_\alpha\Big] \hspace{8pt} , \hspace{8pt} \cdots \hspace{8pt} , \hspace{8pt} \Big[\Big[\Big[\Big[\widetilde{\mathcal{S}}^{(1)}_{\text{graviton},+}(\widehat{p}_s),Q^I_\alpha\Big],Q^J_\beta\Big],Q^K_\gamma\Big]Q^L_\delta\Big]
\end{equation}
are given respectively by:
\begin{align}
    -\frac{\kappa}{4} &\sumop_{m=1}^n \frac{1}{\widehat{p}_s \cdot p_m} (\widehat{p}_s \cdot J_m \cdot \varepsilon_+)(\widehat{p}_s \cdot J_m \cdot \varepsilon_+) \nonumber \\
        \frac{\kappa}{8} &\sumop_{m=1}^n \frac{i\sqrt{2}}{\widehat{p}_s \cdot p_m} \bigg(\big[\widehat{p}_s \hspace{2pt} Q^I_m\big] \hspace{2pt} (\widehat{p}_s \cdot J_m \cdot \varepsilon_+) +(\widehat{p}_s \cdot J_m \cdot \varepsilon_+) \big[\widehat{p}_s \hspace{2pt} Q^I_m \big]\bigg) |\widehat{p}_s]_{\alpha} \nonumber \\
        \frac{\kappa}{8} &\sumop_{m=1}^n \frac{1}{\widehat{p}_s \cdot p_m} \bigg(\big[\widehat{p}_s \hspace{2pt} Q^I_m \big]\big[\widehat{p}_s \hspace{2pt} Q^J_m \big] - \big[\widehat{p}_s \hspace{2pt} Q^J_m\big]\big[\widehat{p}_s \hspace{2pt} Q^I_m\big] - 2 \sqrt{2} i \hspace{2pt}(\widehat{p}_s \cdot J_m \cdot \varepsilon_+) \hspace{2pt} Z_m^{IJ}\bigg)|\widehat{p}_s]_{\alpha} |\widehat{p}_s]_{\beta} \nonumber\\
        -\frac{\kappa}{4} &\sumop_{m=1}^n \frac{1}{\widehat{p}_s \cdot p_m} \bigg(\big[\widehat{p}_s \hspace{2pt} Q^I_m \big] \hspace{2pt} Z_m^{JK} -\big[\widehat{p}_s \hspace{2pt} Q^J_m \big] \hspace{2pt} Z_m^{IK} + \big[\widehat{p}_s \hspace{2pt} Q^K_m \big] \hspace{2pt} Z_m^{IJ}\bigg)|\widehat{p}_s]_{\alpha}|\widehat{p}_s]_{\beta}|\widehat{p}_s]_{\gamma} \nonumber \\
        \frac{\kappa}{4} &\sumop_{m=1}^n \frac{1}{\widehat{p}_s \cdot p_m} \bigg(Z_m^{IJ} Z_m^{KL} -Z_m^{IK} Z_m^{JL} + Z_m^{IL} Z_m^{JK} \bigg) |\widehat{p}_s]_{\alpha}|\widehat{p}_s]_{\beta}|\widehat{p}_s]_{\gamma}|\widehat{p}_s]_{\delta}.
\end{align}
For the negative helicity graviton, the analogous commutators are respectively given by:
\begin{align}
    -\frac{\kappa}{4} &\sumop_{m=1}^n \frac{1}{\widehat{p}_s \cdot p_m} (\widehat{p}_s \cdot J_m \cdot \varepsilon_-)(\widehat{p}_s \cdot J_m \cdot \varepsilon_-) \nonumber \\
        \frac{\kappa}{8} &\sumop_{m=1}^n \frac{i\sqrt{2}}{\widehat{p}_s \cdot p_m} \bigg(\big\langle \overline{Q}^I_m \hspace{2pt} \widehat{p}_s\big\rangle \hspace{2pt} (\widehat{p}_s \cdot J_m \cdot \varepsilon_-) +(\widehat{p}_s \cdot J_m \cdot \varepsilon_-) \big\langle \overline{Q}^I_m \hspace{2pt} \widehat{p}_s \big\rangle\bigg) \langle\widehat{p}_s|_{\dot\alpha} \nonumber \\
        \frac{\kappa}{8} &\sumop_{m=1}^n \frac{1}{\widehat{p}_s \cdot p_m} \bigg(\big\langle \overline{Q}^I_m \hspace{2pt} \widehat{p}_s \big\rangle \big\langle \overline{Q}^J_m \hspace{2pt} \widehat{p}_s \big\rangle - \big\langle \overline{Q}^J_m \hspace{2pt} \widehat{p}_s \big\rangle \big\langle \overline{Q}^I_m \hspace{2pt} \widehat{p}_s\big\rangle - 2 \sqrt{2} i \hspace{2pt}(\widehat{p}_s \cdot J_m \cdot \varepsilon_-) \hspace{2pt} \overline{Z}_m^{IJ}\bigg)\langle\widehat{p}_s|_{\dot \alpha} \langle\widehat{p}_s|_{\dot \beta} \nonumber\\
        -\frac{\kappa}{4} &\sumop_{m=1}^n \frac{1}{\widehat{p}_s \cdot p_m} \bigg(\big\langle \overline{Q}^I_m \hspace{2pt} \widehat{p}_s \big\rangle \hspace{2pt} \overline{Z}_m^{JK} -\big\langle \overline{Q}^J_m \hspace{2pt} \widehat{p}_s \big\rangle \hspace{2pt} \overline{Z}_m^{IK} + \big\langle \overline{Q}^K_m 
        \hspace{2pt} \widehat{p}_s \big\rangle \hspace{2pt} \overline{Z}_m^{IJ}\bigg)\langle\widehat{p}_s|_{\dot \alpha}\langle\widehat{p}_s|_{\dot \beta}\langle\widehat{p}_s|_{\dot \gamma} \nonumber \\
        \frac{\kappa}{4} &\sumop_{m=1}^n \frac{1}{\widehat{p}_s \cdot p_m} \bigg(\overline{Z}_m^{IJ} \overline{Z}_m^{KL} -\overline{Z}_m^{IK} \overline{Z}_m^{JL} + \overline{Z}_m^{IL} \overline{Z}_m^{JK} \bigg) \langle\widehat{p}_s|_{\dot \alpha}\langle\widehat{p}_s|_{\dot \beta}\langle\widehat{p}_s|_{\dot \gamma}\langle\widehat{p}_s|_{\dot \delta}.
\end{align}
Note in all cases that expressions for positive and negative helicity soft theorems are related by CPT conjugation, as they must be.

\section{Comparison with Prior Results}
\label{appendix: comparison with literature}

In this appendix, we use our formalism to reproduce the calculation of the soft photino theorem in $\mathcal{N} = 1$ super QED with a charged, massless chiral multiplet \cite{Dumitrescu:2015fej}. The spectrum of this theory consists of a photon and photino in the vector multiplet and a fermion and complex scalar in a chiral multiplet. The authors choose a normalization where annihilation operators in the chiral multiplet are related through supersymmetry as follows:
\begin{equation}
    \begin{split}
        \Big[a_{\text{fermion},+}(p),Q_{\alpha}\Big] &=  \sqrt{2} i \hspace{2pt} a_{\text{scalar},+}(p) \hspace{2pt}|p]_{\alpha}\hspace{26.5pt} \Big[a_{\text{fermion},-}(p),Q_{\alpha}\Big] = 0\\
        \Big[a_{\text{scalar},+}(p),Q_{\alpha}\Big] &= 0 \hspace{119pt} \Big[a_{\text{scalar},-}(p),Q_{\alpha}\Big] = -\sqrt{2} i \hspace{2pt} a_{\text{fermion},-}(p) \hspace{2pt} |p]_{\alpha},
        \label{eqn: N=1 QED multiplet}
    \end{split}
\end{equation}

The leading positive helicity soft photino theorem computed in \cite{Dumitrescu:2015fej} is given by
\begin{equation}
    \mathcal{S}_{\text{photino},+}(p_s) = \sqrt{2} i e ~ \omega_s^{-1/2} ~ \sumop_{m=1}^n \frac{q_m}{\langle p_m \hspace{2pt} \widehat{p}_s \rangle} ~ \mathcal{F}_m,
    \label{eqn: soft photino strominger}
\end{equation}
where the authors have defined the particle changing operator $\mathcal{F}$ according to the commutators
\begin{equation}
    \begin{split}
        \Big[a_{\text{fermion},+}(p),\mathcal{F}\Big] &= a_{\text{scalar},+}(p) \hspace{70pt} \Big[a_{\text{fermion},-}(p),\mathcal{F}\Big] = 0 \\
        \Big[a_{\text{scalar},+}(p),\mathcal{F}\Big] &= 0 \hspace{122pt} \Big[a_{\text{scalar},-}(p),\mathcal{F}\Big] = - a_{\text{fermion},-}(p).\hspace{40pt} \\
    \end{split}
\end{equation}
It does not matter how $\mathcal{F}$ acts on particles in the vector multiplet as they all have charge $q_m = 0$.

Notice the striking similarity between how the particle changing operator and the supercharge $Q_{\alpha}$ are defined. Indeed, for particles in the chiral multiplet
\begin{equation}
    \Big[a(p),Q_{\alpha}\Big] = \sqrt{2}i \hspace{2pt} |p]_{\alpha} \hspace{2pt} \Big[a(p),\mathcal{F}\Big].
\end{equation}
This expression allows us to translate between the leading soft photino theorem derived in this work and Equation \eqref{eqn: soft photino strominger} verifying that they are in agreement\footnote{We note that the soft photino theorem in this example is uncorrected by the theory-dependent operators $\mathcal{F}^{I_1 \cdots I_n}$ because the dimension-five operators which generate them are absent from the Lagrangian we are studying.}
\begin{equation}
    \begin{split}
        \mathcal{S}_{\text{photino},+}(p_s) &= -\frac{e}{2} \hspace{2pt} \omega_s^{-1/2} \sumop_{m=1}^n \frac{q_m}{\widehat{p}_s \cdot p_m} \big[\hspace{1pt} \widehat{p}_s \hspace{2pt} Q_m \big]\\
        &= -\frac{e}{2} \hspace{2pt} \omega_s^{-1/2} \sumop_{m=1}^n \frac{2 \hspace{1pt}q_m}{\langle p_m \hspace{2pt} \widehat{p}_s \rangle \hspace{1pt} [p_m \hspace{2pt} \widehat{p}_s]} \sqrt{2} i \hspace{2pt} \mathcal{F}_m \hspace{2pt} \big[\hspace{1pt} \widehat{p}_s \hspace{2pt} p_m\big] \\
        &= \sqrt{2} i e ~ \omega_s^{-1/2} ~ \sumop_{m=1}^n \frac{q_m}{\langle p_m \hspace{2pt} \widehat{p}_s \rangle} ~ \mathcal{F}_m.
    \end{split}
    \label{eqn: soft photino tropper}
\end{equation}

\bibliography{mybib.bib}

\begin{thebibliography}{10}%
\makeatletter
\providecommand \@ifxundefined [1]{%
 \ifx #1\undefined \expandafter \@firstoftwo
 \else \expandafter \@secondoftwo
\fi
}%
\providecommand \@ifnum [1]{%
 \ifnum #1\expandafter \@firstoftwo
 \else \expandafter \@secondoftwo
\fi
}%
\providecommand \enquote [1]{``#1''}%
\providecommand \bibnamefont  [1]{#1}%
\providecommand \bibfnamefont [1]{#1}%
\providecommand \citenamefont [1]{#1}%
\providecommand\href[0]{\@sanitize\@href}%
\providecommand\@href[1]{\endgroup\@@startlink{#1}\endgroup\@@href}%
\providecommand\@@href[1]{#1\@@endlink}%
\providecommand \@sanitize [0]{\begingroup\catcode`\&12\catcode`\#12\relax}%
\@ifxundefined \pdfoutput {\@firstoftwo}{%
 \@ifnum{\z@=\pdfoutput}{\@firstoftwo}{\@secondoftwo}%
}{%
 \providecommand\@@startlink[1]{\leavevmode\special{html:<a href="#1">}}%
 \providecommand\@@endlink[0]{\special{html:</a>}}%
}{%
 \providecommand\@@startlink[1]{%
  \leavevmode
  \pdfstartlink
   attr{/Border[0 0 1 ]/H/I/C[0 1 1]}%
   user{/Subtype/Link/A<</Type/Action/S/URI/URI(#1)>>}%
  \relax
 }%
 \providecommand\@@endlink[0]{\pdfendlink}%
}%
\providecommand \url  [0]{\begingroup\@sanitize \@url }%
\providecommand \@url [1]{\endgroup\@href {#1}{\urlprefix}}%
\providecommand \urlprefix [0]{URL }%
\providecommand \Eprint[0]{\href }%
\@ifxundefined \urlstyle {%
  \providecommand \doi [1]{doi:\discretionary{}{}{}#1}%
}{%
  \providecommand \doi [0]{doi:\discretionary{}{}{}\begingroup
  \urlstyle{rm}\Url }%
}%
\providecommand \doibase [0]{http://dx.doi.org/}%
\providecommand \Doi[1]{\href{\doibase#1}}%
\providecommand \bibAnnote [3]{%
  \BibitemShut{#1}%
  \begin{quotation}\noindent
    \textsc{Key:}\ #2\\\textsc{Annotation:}\ #3%
  \end{quotation}%
}%
\providecommand \bibAnnoteFile [2]{%
  \IfFileExists{#2}{\bibAnnote {#1} {#2} {\input{#2}}}{}%
}%
\providecommand \typeout [0]{\immediate \write \m@ne }%
\providecommand \selectlanguage [0]{\@gobble}%
\providecommand \bibinfo [0]{\@secondoftwo}%
\providecommand \bibfield [0]{\@secondoftwo}%
\providecommand \translation [1]{[#1]}%
\providecommand \BibitemOpen[0]{}%
\providecommand \bibitemStop [0]{}%
\providecommand \bibitemNoStop [0]{.\EOS\space}%
\providecommand \EOS [0]{\spacefactor3000\relax}%
\providecommand \BibitemShut [1]{\csname bibitem#1\endcsname}%
\bibitem{Weinberg:1965nx}%
  \BibitemOpen
  \bibfield{author}{%
  \bibinfo {author} {\bibfnamefont{Steven}\ \bibnamefont{Weinberg}},\ }%
  \bibfield{title}{%
  \enquote{\bibinfo {title} {{Infrared photons and gravitons}},}\ }%
  \bibfield{journal}{%
  \Doi{10.1103/PhysRev.140.B516}{\bibinfo {journal} {Phys. Rev.}}\ }%
  \textbf{\bibinfo {volume} {140}},\ \bibinfo {pages} {B516--B524} (\bibinfo
  {year} {1965})%
  \bibAnnoteFile{NoStop}{Weinberg:1965nx}%
\bibitem{Low:1958sn}%
  \BibitemOpen
  \bibfield{author}{%
  \bibinfo {author} {\bibfnamefont{F.~E.}\ \bibnamefont{Low}},\ }%
  \bibfield{title}{%
  \enquote{\bibinfo {title} {{Bremsstrahlung of very low-energy quanta in
  elementary particle collisions}},}\ }%
  \bibfield{journal}{%
  \Doi{10.1103/PhysRev.110.974}{\bibinfo {journal} {Phys. Rev.}}\ }%
  \textbf{\bibinfo {volume} {110}},\ \bibinfo {pages} {974--977} (\bibinfo
  {year} {1958})%
  \bibAnnoteFile{NoStop}{Low:1958sn}%
\bibitem{Cheung:2014dqa}%
  \BibitemOpen
  \bibfield{author}{%
  \bibinfo {author} {\bibfnamefont{Clifford}\ \bibnamefont{Cheung}}, \bibinfo
  {author} {\bibfnamefont{Karol}\ \bibnamefont{Kampf}}, \bibinfo {author}
  {\bibfnamefont{Jiri}\ \bibnamefont{Novotny}},\ and\ \bibinfo {author}
  {\bibfnamefont{Jaroslav}\ \bibnamefont{Trnka}},\ }%
  \bibfield{title}{%
  \enquote{\bibinfo {title} {{Effective Field Theories from Soft Limits of
  Scattering Amplitudes}},}\ }%
  \bibfield{journal}{%
  \Doi{10.1103/PhysRevLett.114.221602}{\bibinfo {journal} {Phys. Rev. Lett.}}\
  }%
  \textbf{\bibinfo {volume} {114}},\ \bibinfo {pages} {221602} (\bibinfo {year}
  {2015}),\ \Eprint{http://arxiv.org/abs/1412.4095}{arXiv:1412.4095 [hep-th]}%
  \bibAnnoteFile{NoStop}{Cheung:2014dqa}%
\bibitem{Cheung:2015ota}%
  \BibitemOpen
  \bibfield{author}{%
  \bibinfo {author} {\bibfnamefont{Clifford}\ \bibnamefont{Cheung}}, \bibinfo
  {author} {\bibfnamefont{Karol}\ \bibnamefont{Kampf}}, \bibinfo {author}
  {\bibfnamefont{Jiri}\ \bibnamefont{Novotny}}, \bibinfo {author}
  {\bibfnamefont{Chia-Hsien}\ \bibnamefont{Shen}},\ and\ \bibinfo {author}
  {\bibfnamefont{Jaroslav}\ \bibnamefont{Trnka}},\ }%
  \bibfield{title}{%
  \enquote{\bibinfo {title} {{On-Shell Recursion Relations for Effective Field
  Theories}},}\ }%
  \bibfield{journal}{%
  \Doi{10.1103/PhysRevLett.116.041601}{\bibinfo {journal} {Phys. Rev. Lett.}}\
  }%
  \textbf{\bibinfo {volume} {116}},\ \bibinfo {pages} {041601} (\bibinfo {year}
  {2016}),\ \Eprint{http://arxiv.org/abs/1509.03309}{arXiv:1509.03309
  [hep-th]}%
  \bibAnnoteFile{NoStop}{Cheung:2015ota}%
\bibitem{Cheung:2016drk}%
  \BibitemOpen
  \bibfield{author}{%
  \bibinfo {author} {\bibfnamefont{Clifford}\ \bibnamefont{Cheung}}, \bibinfo
  {author} {\bibfnamefont{Karol}\ \bibnamefont{Kampf}}, \bibinfo {author}
  {\bibfnamefont{Jiri}\ \bibnamefont{Novotny}}, \bibinfo {author}
  {\bibfnamefont{Chia-Hsien}\ \bibnamefont{Shen}},\ and\ \bibinfo {author}
  {\bibfnamefont{Jaroslav}\ \bibnamefont{Trnka}},\ }%
  \bibfield{title}{%
  \enquote{\bibinfo {title} {{A Periodic Table of Effective Field Theories}},}\
  }%
  \bibfield{journal}{%
  \Doi{10.1007/JHEP02(2017)020}{\bibinfo {journal} {JHEP}}\ }%
  \textbf{\bibinfo {volume} {02}},\ \bibinfo {pages} {020} (\bibinfo {year}
  {2017}),\ \Eprint{http://arxiv.org/abs/1611.03137}{arXiv:1611.03137
  [hep-th]}%
  \bibAnnoteFile{NoStop}{Cheung:2016drk}%
\bibitem{Zhou:2022orv}%
  \BibitemOpen
  \bibfield{author}{%
  \bibinfo {author} {\bibfnamefont{Kang}\ \bibnamefont{Zhou}},\ }%
  \bibfield{title}{%
  \enquote{\bibinfo {title} {{Tree level amplitudes from soft theorems}},}\ }%
  \bibfield{journal}{%
  \Doi{10.1007/JHEP03(2023)021}{\bibinfo {journal} {JHEP}}\ }%
  \textbf{\bibinfo {volume} {03}},\ \bibinfo {pages} {021} (\bibinfo {year}
  {2023}),\ \Eprint{http://arxiv.org/abs/2212.12892}{arXiv:2212.12892
  [hep-th]}%
  \bibAnnoteFile{NoStop}{Zhou:2022orv}%
\bibitem{Kalyanapuram:2020epb}%
  \BibitemOpen
  \bibfield{author}{%
  \bibinfo {author} {\bibfnamefont{Nikhil}\ \bibnamefont{Kalyanapuram}},\ }%
  \bibfield{title}{%
  \enquote{\bibinfo {title} {{Soft Gravity by Squaring Soft QED on the
  Celestial Sphere}},}\ }%
  \bibfield{journal}{%
  \Doi{10.1103/PhysRevD.103.085016}{\bibinfo {journal} {Phys. Rev. D}}\ }%
  \textbf{\bibinfo {volume} {103}},\ \bibinfo {pages} {085016} (\bibinfo {year}
  {2021}),\ \Eprint{http://arxiv.org/abs/2011.11412}{arXiv:2011.11412
  [hep-th]}%
  \bibAnnoteFile{NoStop}{Kalyanapuram:2020epb}%
\bibitem{Bautista:2019tdr}%
  \BibitemOpen
  \bibfield{author}{%
  \bibinfo {author} {\bibfnamefont{Yilber~Fabian}\ \bibnamefont{Bautista}}\
  and\ \bibinfo {author} {\bibfnamefont{Alfredo}\ \bibnamefont{Guevara}},\ }%
  \bibfield{title}{%
  \enquote{\bibinfo {title} {{From Scattering Amplitudes to Classical Physics:
  Universality, Double Copy and Soft Theorems}},}\ }%
   (\bibinfo {month} {3}\ \bibinfo {year} {2019}),\
  \Eprint{http://arxiv.org/abs/1903.12419}{arXiv:1903.12419 [hep-th]}%
  \bibAnnoteFile{NoStop}{Bautista:2019tdr}%
\bibitem{Strominger:2017zoo}%
  \BibitemOpen
  \bibfield{author}{%
  \bibinfo {author} {\bibfnamefont{Andrew}\ \bibnamefont{Strominger}},\ }%
  \bibfield{title}{%
  \enquote{\bibinfo {title} {{Lectures on the Infrared Structure of Gravity and
  Gauge Theory}},}\ }%
   (\bibinfo {month} {3}\ \bibinfo {year} {2017}),\
  \Eprint{http://arxiv.org/abs/1703.05448}{arXiv:1703.05448 [hep-th]}%
  \bibAnnoteFile{NoStop}{Strominger:2017zoo}%
\bibitem{Strominger:2013jfa}%
  \BibitemOpen
  \bibfield{author}{%
  \bibinfo {author} {\bibfnamefont{Andrew}\ \bibnamefont{Strominger}},\ }%
  \bibfield{title}{%
  \enquote{\bibinfo {title} {{On BMS Invariance of Gravitational
  Scattering}},}\ }%
  \bibfield{journal}{%
  \Doi{10.1007/JHEP07(2014)152}{\bibinfo {journal} {JHEP}}\ }%
  \textbf{\bibinfo {volume} {07}},\ \bibinfo {pages} {152} (\bibinfo {year}
  {2014}),\ \Eprint{http://arxiv.org/abs/1312.2229}{arXiv:1312.2229 [hep-th]}%
  \bibAnnoteFile{NoStop}{Strominger:2013jfa}%
\bibitem{Strominger:2013lka}%
  \BibitemOpen
  \bibfield{author}{%
  \bibinfo {author} {\bibfnamefont{Andrew}\ \bibnamefont{Strominger}},\ }%
  \bibfield{title}{%
  \enquote{\bibinfo {title} {{Asymptotic Symmetries of Yang-Mills Theory}},}\
  }%
  \bibfield{journal}{%
  \Doi{10.1007/JHEP07(2014)151}{\bibinfo {journal} {JHEP}}\ }%
  \textbf{\bibinfo {volume} {07}},\ \bibinfo {pages} {151} (\bibinfo {year}
  {2014}),\ \Eprint{http://arxiv.org/abs/1308.0589}{arXiv:1308.0589 [hep-th]}%
  \bibAnnoteFile{NoStop}{Strominger:2013lka}%
\bibitem{Lysov:2014csa}%
  \BibitemOpen
  \bibfield{author}{%
  \bibinfo {author} {\bibfnamefont{Vyacheslav}\ \bibnamefont{Lysov}}, \bibinfo
  {author} {\bibfnamefont{Sabrina}\ \bibnamefont{Pasterski}},\ and\ \bibinfo
  {author} {\bibfnamefont{Andrew}\ \bibnamefont{Strominger}},\ }%
  \bibfield{title}{%
  \enquote{\bibinfo {title} {{Low\textquoteright{}s Subleading Soft Theorem as
  a Symmetry of QED}},}\ }%
  \bibfield{journal}{%
  \Doi{10.1103/PhysRevLett.113.111601}{\bibinfo {journal} {Phys. Rev. Lett.}}\
  }%
  \textbf{\bibinfo {volume} {113}},\ \bibinfo {pages} {111601} (\bibinfo {year}
  {2014}),\ \Eprint{http://arxiv.org/abs/1407.3814}{arXiv:1407.3814 [hep-th]}%
  \bibAnnoteFile{NoStop}{Lysov:2014csa}%
\bibitem{He:2014laa}%
  \BibitemOpen
  \bibfield{author}{%
  \bibinfo {author} {\bibfnamefont{Temple}\ \bibnamefont{He}}, \bibinfo
  {author} {\bibfnamefont{Vyacheslav}\ \bibnamefont{Lysov}}, \bibinfo {author}
  {\bibfnamefont{Prahar}\ \bibnamefont{Mitra}},\ and\ \bibinfo {author}
  {\bibfnamefont{Andrew}\ \bibnamefont{Strominger}},\ }%
  \bibfield{title}{%
  \enquote{\bibinfo {title} {{BMS supertranslations and
  Weinberg\textquoteright{}s soft graviton theorem}},}\ }%
  \bibfield{journal}{%
  \Doi{10.1007/JHEP05(2015)151}{\bibinfo {journal} {JHEP}}\ }%
  \textbf{\bibinfo {volume} {05}},\ \bibinfo {pages} {151} (\bibinfo {year}
  {2015}),\ \Eprint{http://arxiv.org/abs/1401.7026}{arXiv:1401.7026 [hep-th]}%
  \bibAnnoteFile{NoStop}{He:2014laa}%
\bibitem{He:2014cra}%
  \BibitemOpen
  \bibfield{author}{%
  \bibinfo {author} {\bibfnamefont{Temple}\ \bibnamefont{He}}, \bibinfo
  {author} {\bibfnamefont{Prahar}\ \bibnamefont{Mitra}}, \bibinfo {author}
  {\bibfnamefont{Achilleas~P.}\ \bibnamefont{Porfyriadis}},\ and\ \bibinfo
  {author} {\bibfnamefont{Andrew}\ \bibnamefont{Strominger}},\ }%
  \bibfield{title}{%
  \enquote{\bibinfo {title} {{New Symmetries of Massless QED}},}\ }%
  \bibfield{journal}{%
  \Doi{10.1007/JHEP10(2014)112}{\bibinfo {journal} {JHEP}}\ }%
  \textbf{\bibinfo {volume} {10}},\ \bibinfo {pages} {112} (\bibinfo {year}
  {2014}),\ \Eprint{http://arxiv.org/abs/1407.3789}{arXiv:1407.3789 [hep-th]}%
  \bibAnnoteFile{NoStop}{He:2014cra}%
\bibitem{Miller:2021hty}%
  \BibitemOpen
  \bibfield{author}{%
  \bibinfo {author} {\bibfnamefont{Noah}\ \bibnamefont{Miller}},\ }%
  \bibfield{title}{%
  \enquote{\bibinfo {title} {{From Noether's Theorem to Bremsstrahlung: a
  pedagogical introduction to large gauge transformations and classical soft
  theorems}},}\ }%
   (\bibinfo {month} {12}\ \bibinfo {year} {2021}),\
  \Eprint{http://arxiv.org/abs/2112.05289}{arXiv:2112.05289 [hep-th]}%
  \bibAnnoteFile{NoStop}{Miller:2021hty}%
\bibitem{Strominger:2014pwa}%
  \BibitemOpen
  \bibfield{author}{%
  \bibinfo {author} {\bibfnamefont{Andrew}\ \bibnamefont{Strominger}}\ and\
  \bibinfo {author} {\bibfnamefont{Alexander}\ \bibnamefont{Zhiboedov}},\ }%
  \bibfield{title}{%
  \enquote{\bibinfo {title} {{Gravitational Memory, BMS Supertranslations and
  Soft Theorems}},}\ }%
  \bibfield{journal}{%
  \Doi{10.1007/JHEP01(2016)086}{\bibinfo {journal} {JHEP}}\ }%
  \textbf{\bibinfo {volume} {01}},\ \bibinfo {pages} {086} (\bibinfo {year}
  {2016}),\ \Eprint{http://arxiv.org/abs/1411.5745}{arXiv:1411.5745 [hep-th]}%
  \bibAnnoteFile{NoStop}{Strominger:2014pwa}%
\bibitem{Pasterski:2015tva}%
  \BibitemOpen
  \bibfield{author}{%
  \bibinfo {author} {\bibfnamefont{Sabrina}\ \bibnamefont{Pasterski}}, \bibinfo
  {author} {\bibfnamefont{Andrew}\ \bibnamefont{Strominger}},\ and\ \bibinfo
  {author} {\bibfnamefont{Alexander}\ \bibnamefont{Zhiboedov}},\ }%
  \bibfield{title}{%
  \enquote{\bibinfo {title} {{New Gravitational Memories}},}\ }%
  \bibfield{journal}{%
  \Doi{10.1007/JHEP12(2016)053}{\bibinfo {journal} {JHEP}}\ }%
  \textbf{\bibinfo {volume} {12}},\ \bibinfo {pages} {053} (\bibinfo {year}
  {2016}),\ \Eprint{http://arxiv.org/abs/1502.06120}{arXiv:1502.06120
  [hep-th]}%
  \bibAnnoteFile{NoStop}{Pasterski:2015tva}%
\bibitem{Pate:2017vwa}%
  \BibitemOpen
  \bibfield{author}{%
  \bibinfo {author} {\bibfnamefont{Monica}\ \bibnamefont{Pate}}, \bibinfo
  {author} {\bibfnamefont{Ana-Maria}\ \bibnamefont{Raclariu}},\ and\ \bibinfo
  {author} {\bibfnamefont{Andrew}\ \bibnamefont{Strominger}},\ }%
  \bibfield{title}{%
  \enquote{\bibinfo {title} {{Color Memory: A Yang-Mills Analog of
  Gravitational Wave Memory}},}\ }%
  \bibfield{journal}{%
  \Doi{10.1103/PhysRevLett.119.261602}{\bibinfo {journal} {Phys. Rev. Lett.}}\
  }%
  \textbf{\bibinfo {volume} {119}},\ \bibinfo {pages} {261602} (\bibinfo {year}
  {2017}),\ \Eprint{http://arxiv.org/abs/1707.08016}{arXiv:1707.08016
  [hep-th]}%
  \bibAnnoteFile{NoStop}{Pate:2017vwa}%
\bibitem{Chatterjee:2017zeb}%
  \BibitemOpen
  \bibfield{author}{%
  \bibinfo {author} {\bibfnamefont{Atreya}\ \bibnamefont{Chatterjee}}\ and\
  \bibinfo {author} {\bibfnamefont{David~A.}\ \bibnamefont{Lowe}},\ }%
  \bibfield{title}{%
  \enquote{\bibinfo {title} {{BMS symmetry, soft particles and memory}},}\ }%
  \bibfield{journal}{%
  \Doi{10.1088/1361-6382/aab5cc}{\bibinfo {journal} {Class. Quant. Grav.}}\ }%
  \textbf{\bibinfo {volume} {35}},\ \bibinfo {pages} {094001} (\bibinfo {year}
  {2018}),\ \Eprint{http://arxiv.org/abs/1712.03211}{arXiv:1712.03211
  [hep-th]}%
  \bibAnnoteFile{NoStop}{Chatterjee:2017zeb}%
\bibitem{Laddha:2018vbn}%
  \BibitemOpen
  \bibfield{author}{%
  \bibinfo {author} {\bibfnamefont{Alok}\ \bibnamefont{Laddha}}\ and\ \bibinfo
  {author} {\bibfnamefont{Ashoke}\ \bibnamefont{Sen}},\ }%
  \bibfield{title}{%
  \enquote{\bibinfo {title} {{Observational Signature of the Logarithmic Terms
  in the Soft Graviton Theorem}},}\ }%
  \bibfield{journal}{%
  \Doi{10.1103/PhysRevD.100.024009}{\bibinfo {journal} {Phys. Rev. D}}\ }%
  \textbf{\bibinfo {volume} {100}},\ \bibinfo {pages} {024009} (\bibinfo {year}
  {2019}),\ \Eprint{http://arxiv.org/abs/1806.01872}{arXiv:1806.01872
  [hep-th]}%
  \bibAnnoteFile{NoStop}{Laddha:2018vbn}%
\bibitem{Ball:2018prg}%
  \BibitemOpen
  \bibfield{author}{%
  \bibinfo {author} {\bibfnamefont{Adam}\ \bibnamefont{Ball}}, \bibinfo
  {author} {\bibfnamefont{Monica}\ \bibnamefont{Pate}}, \bibinfo {author}
  {\bibfnamefont{Ana-Maria}\ \bibnamefont{Raclariu}}, \bibinfo {author}
  {\bibfnamefont{Andrew}\ \bibnamefont{Strominger}},\ and\ \bibinfo {author}
  {\bibfnamefont{Raju}\ \bibnamefont{Venugopalan}},\ }%
  \bibfield{title}{%
  \enquote{\bibinfo {title} {{Measuring color memory in a color glass
  condensate at electron\textendash{}ion colliders}},}\ }%
  \bibfield{journal}{%
  \Doi{10.1016/j.aop.2019.04.010}{\bibinfo {journal} {Annals Phys.}}\ }%
  \textbf{\bibinfo {volume} {407}},\ \bibinfo {pages} {15--28} (\bibinfo {year}
  {2019}),\ \Eprint{http://arxiv.org/abs/1805.12224}{arXiv:1805.12224
  [hep-ph]}%
  \bibAnnoteFile{NoStop}{Ball:2018prg}%
\bibitem{Raclariu:2021zjz}%
  \BibitemOpen
  \bibfield{author}{%
  \bibinfo {author} {\bibfnamefont{Ana-Maria}\ \bibnamefont{Raclariu}},\ }%
  \bibfield{title}{%
  \enquote{\bibinfo {title} {{Lectures on Celestial Holography}},}\ }%
   (\bibinfo {month} {7}\ \bibinfo {year} {2021}),\
  \Eprint{http://arxiv.org/abs/2107.02075}{arXiv:2107.02075 [hep-th]}%
  \bibAnnoteFile{NoStop}{Raclariu:2021zjz}%
\bibitem{Pasterski:2021rjz}%
  \BibitemOpen
  \bibfield{author}{%
  \bibinfo {author} {\bibfnamefont{Sabrina}\ \bibnamefont{Pasterski}},\ }%
  \bibfield{title}{%
  \enquote{\bibinfo {title} {{Lectures on celestial amplitudes}},}\ }%
  \bibfield{journal}{%
  \Doi{10.1140/epjc/s10052-021-09846-7}{\bibinfo {journal} {Eur. Phys. J. C}}\
  }%
  \textbf{\bibinfo {volume} {81}},\ \bibinfo {pages} {1062} (\bibinfo {year}
  {2021}),\ \Eprint{http://arxiv.org/abs/2108.04801}{arXiv:2108.04801
  [hep-th]}%
  \bibAnnoteFile{NoStop}{Pasterski:2021rjz}%
\bibitem{Pasterski:2023ikd}%
  \BibitemOpen
  \bibfield{author}{%
  \bibinfo {author} {\bibfnamefont{Sabrina}\ \bibnamefont{Pasterski}},\ }%
  \bibfield{title}{%
  \enquote{\bibinfo {title} {{A Chapter on Celestial Holography}},}\ }%
   (\bibinfo {month} {10}\ \bibinfo {year} {2023}),\
  \Eprint{http://arxiv.org/abs/2310.04932}{arXiv:2310.04932 [hep-th]}%
  \bibAnnoteFile{NoStop}{Pasterski:2023ikd}%
\bibitem{Kapec:2016jld}%
  \BibitemOpen
  \bibfield{author}{%
  \bibinfo {author} {\bibfnamefont{Daniel}\ \bibnamefont{Kapec}}, \bibinfo
  {author} {\bibfnamefont{Prahar}\ \bibnamefont{Mitra}}, \bibinfo {author}
  {\bibfnamefont{Ana-Maria}\ \bibnamefont{Raclariu}},\ and\ \bibinfo {author}
  {\bibfnamefont{Andrew}\ \bibnamefont{Strominger}},\ }%
  \bibfield{title}{%
  \enquote{\bibinfo {title} {{2D Stress Tensor for 4D Gravity}},}\ }%
  \bibfield{journal}{%
  \Doi{10.1103/PhysRevLett.119.121601}{\bibinfo {journal} {Phys. Rev. Lett.}}\
  }%
  \textbf{\bibinfo {volume} {119}},\ \bibinfo {pages} {121601} (\bibinfo {year}
  {2017}),\ \Eprint{http://arxiv.org/abs/1609.00282}{arXiv:1609.00282
  [hep-th]}%
  \bibAnnoteFile{NoStop}{Kapec:2016jld}%
\bibitem{Nande:2017dba}%
  \BibitemOpen
  \bibfield{author}{%
  \bibinfo {author} {\bibfnamefont{Anjalika}\ \bibnamefont{Nande}}, \bibinfo
  {author} {\bibfnamefont{Monica}\ \bibnamefont{Pate}},\ and\ \bibinfo {author}
  {\bibfnamefont{Andrew}\ \bibnamefont{Strominger}},\ }%
  \bibfield{title}{%
  \enquote{\bibinfo {title} {{Soft Factorization in QED from 2D Kac-Moody
  Symmetry}},}\ }%
  \bibfield{journal}{%
  \Doi{10.1007/JHEP02(2018)079}{\bibinfo {journal} {JHEP}}\ }%
  \textbf{\bibinfo {volume} {02}},\ \bibinfo {pages} {079} (\bibinfo {year}
  {2018}),\ \Eprint{http://arxiv.org/abs/1705.00608}{arXiv:1705.00608
  [hep-th]}%
  \bibAnnoteFile{NoStop}{Nande:2017dba}%
\bibitem{Himwich:2020rro}%
  \BibitemOpen
  \bibfield{author}{%
  \bibinfo {author} {\bibfnamefont{Elizabeth}\ \bibnamefont{Himwich}}, \bibinfo
  {author} {\bibfnamefont{Sruthi~A.}\ \bibnamefont{Narayanan}}, \bibinfo
  {author} {\bibfnamefont{Monica}\ \bibnamefont{Pate}}, \bibinfo {author}
  {\bibfnamefont{Nisarga}\ \bibnamefont{Paul}},\ and\ \bibinfo {author}
  {\bibfnamefont{Andrew}\ \bibnamefont{Strominger}},\ }%
  \bibfield{title}{%
  \enquote{\bibinfo {title} {{The Soft $\mathcal{S}$-Matrix in Gravity}},}\ }%
  \bibfield{journal}{%
  \Doi{10.1007/JHEP09(2020)129}{\bibinfo {journal} {JHEP}}\ }%
  \textbf{\bibinfo {volume} {09}},\ \bibinfo {pages} {129} (\bibinfo {year}
  {2020}),\ \Eprint{http://arxiv.org/abs/2005.13433}{arXiv:2005.13433
  [hep-th]}%
  \bibAnnoteFile{NoStop}{Himwich:2020rro}%
\bibitem{Strominger:2021lvk}%
  \BibitemOpen
  \bibfield{author}{%
  \bibinfo {author} {\bibfnamefont{Andrew}\ \bibnamefont{Strominger}},\ }%
  \bibfield{title}{%
  \enquote{\bibinfo {title} {{w(1+infinity) and the Celestial Sphere}},}\ }%
   (\bibinfo {month} {5}\ \bibinfo {year} {2021}),\
  \Eprint{http://arxiv.org/abs/2105.14346}{arXiv:2105.14346 [hep-th]}%
  \bibAnnoteFile{NoStop}{Strominger:2021lvk}%
\bibitem{Donnay:2022hkf}%
  \BibitemOpen
  \bibfield{author}{%
  \bibinfo {author} {\bibfnamefont{Laura}\ \bibnamefont{Donnay}}, \bibinfo
  {author} {\bibfnamefont{Kevin}\ \bibnamefont{Nguyen}},\ and\ \bibinfo
  {author} {\bibfnamefont{Romain}\ \bibnamefont{Ruzziconi}},\ }%
  \bibfield{title}{%
  \enquote{\bibinfo {title} {{Loop-corrected subleading soft theorem and the
  celestial stress tensor}},}\ }%
  \bibfield{journal}{%
  \Doi{10.1007/JHEP09(2022)063}{\bibinfo {journal} {JHEP}}\ }%
  \textbf{\bibinfo {volume} {09}},\ \bibinfo {pages} {063} (\bibinfo {year}
  {2022}),\ \Eprint{http://arxiv.org/abs/2205.11477}{arXiv:2205.11477
  [hep-th]}%
  \bibAnnoteFile{NoStop}{Donnay:2022hkf}%
\bibitem{Cachazo:2014fwa}%
  \BibitemOpen
  \bibfield{author}{%
  \bibinfo {author} {\bibfnamefont{Freddy}\ \bibnamefont{Cachazo}}\ and\
  \bibinfo {author} {\bibfnamefont{Andrew}\ \bibnamefont{Strominger}},\ }%
  \bibfield{title}{%
  \enquote{\bibinfo {title} {{Evidence for a New Soft Graviton Theorem}},}\ }%
   (\bibinfo {month} {4}\ \bibinfo {year} {2014}),\
  \Eprint{http://arxiv.org/abs/1404.4091}{arXiv:1404.4091 [hep-th]}%
  \bibAnnoteFile{NoStop}{Cachazo:2014fwa}%
\bibitem{Strominger:2015bla}%
  \BibitemOpen
  \bibfield{author}{%
  \bibinfo {author} {\bibfnamefont{Andrew}\ \bibnamefont{Strominger}},\ }%
  \bibfield{title}{%
  \enquote{\bibinfo {title} {{Magnetic Corrections to the Soft Photon
  Theorem}},}\ }%
  \bibfield{journal}{%
  \Doi{10.1103/PhysRevLett.116.031602}{\bibinfo {journal} {Phys. Rev. Lett.}}\
  }%
  \textbf{\bibinfo {volume} {116}},\ \bibinfo {pages} {031602} (\bibinfo {year}
  {2016}),\ \Eprint{http://arxiv.org/abs/1509.00543}{arXiv:1509.00543
  [hep-th]}%
  \bibAnnoteFile{NoStop}{Strominger:2015bla}%
\bibitem{Hawking:2016msc}%
  \BibitemOpen
  \bibfield{author}{%
  \bibinfo {author} {\bibfnamefont{Stephen~W.}\ \bibnamefont{Hawking}},
  \bibinfo {author} {\bibfnamefont{Malcolm~J.}\ \bibnamefont{Perry}},\ and\
  \bibinfo {author} {\bibfnamefont{Andrew}\ \bibnamefont{Strominger}},\ }%
  \bibfield{title}{%
  \enquote{\bibinfo {title} {{Soft Hair on Black Holes}},}\ }%
  \bibfield{journal}{%
  \Doi{10.1103/PhysRevLett.116.231301}{\bibinfo {journal} {Phys. Rev. Lett.}}\
  }%
  \textbf{\bibinfo {volume} {116}},\ \bibinfo {pages} {231301} (\bibinfo {year}
  {2016}),\ \Eprint{http://arxiv.org/abs/1601.00921}{arXiv:1601.00921
  [hep-th]}%
  \bibAnnoteFile{NoStop}{Hawking:2016msc}%
\bibitem{Hawking:2016sgy}%
  \BibitemOpen
  \bibfield{author}{%
  \bibinfo {author} {\bibfnamefont{Stephen~W.}\ \bibnamefont{Hawking}},
  \bibinfo {author} {\bibfnamefont{Malcolm~J.}\ \bibnamefont{Perry}},\ and\
  \bibinfo {author} {\bibfnamefont{Andrew}\ \bibnamefont{Strominger}},\ }%
  \bibfield{title}{%
  \enquote{\bibinfo {title} {{Superrotation Charge and Supertranslation Hair on
  Black Holes}},}\ }%
  \bibfield{journal}{%
  \Doi{10.1007/JHEP05(2017)161}{\bibinfo {journal} {JHEP}}\ }%
  \textbf{\bibinfo {volume} {05}},\ \bibinfo {pages} {161} (\bibinfo {year}
  {2017}),\ \Eprint{http://arxiv.org/abs/1611.09175}{arXiv:1611.09175
  [hep-th]}%
  \bibAnnoteFile{NoStop}{Hawking:2016sgy}%
\bibitem{Fernandes:2020tsq}%
  \BibitemOpen
  \bibfield{author}{%
  \bibinfo {author} {\bibfnamefont{Karan}\ \bibnamefont{Fernandes}}\ and\
  \bibinfo {author} {\bibfnamefont{Arpita}\ \bibnamefont{Mitra}},\ }%
  \bibfield{title}{%
  \enquote{\bibinfo {title} {{Soft factors from classical scattering on the
  Reissner-Nordstr\"om spacetime}},}\ }%
  \bibfield{journal}{%
  \Doi{10.1103/PhysRevD.102.105015}{\bibinfo {journal} {Phys. Rev. D}}\ }%
  \textbf{\bibinfo {volume} {102}},\ \bibinfo {pages} {105015} (\bibinfo {year}
  {2020}),\ \Eprint{http://arxiv.org/abs/2005.03613}{arXiv:2005.03613
  [hep-th]}%
  \bibAnnoteFile{NoStop}{Fernandes:2020tsq}%
\bibitem{Marotta:2019cip}%
  \BibitemOpen
  \bibfield{author}{%
  \bibinfo {author} {\bibfnamefont{Raffaele}\ \bibnamefont{Marotta}}\ and\
  \bibinfo {author} {\bibfnamefont{Mritunjay}\ \bibnamefont{Verma}},\ }%
  \bibfield{title}{%
  \enquote{\bibinfo {title} {{Soft Theorems from Compactification}},}\ }%
  \bibfield{journal}{%
  \Doi{10.1007/JHEP02(2020)008}{\bibinfo {journal} {JHEP}}\ }%
  \textbf{\bibinfo {volume} {02}},\ \bibinfo {pages} {008} (\bibinfo {year}
  {2020}),\ \Eprint{http://arxiv.org/abs/1911.05099}{arXiv:1911.05099
  [hep-th]}%
  \bibAnnoteFile{NoStop}{Marotta:2019cip}%
\bibitem{Miller:2022fvc}%
  \BibitemOpen
  \bibfield{author}{%
  \bibinfo {author} {\bibfnamefont{Noah}\ \bibnamefont{Miller}}, \bibinfo
  {author} {\bibfnamefont{Andrew}\ \bibnamefont{Strominger}}, \bibinfo {author}
  {\bibfnamefont{Adam}\ \bibnamefont{Tropper}},\ and\ \bibinfo {author}
  {\bibfnamefont{Tianli}\ \bibnamefont{Wang}},\ }%
  \bibfield{title}{%
  \enquote{\bibinfo {title} {{Soft Gravitons in the BFSS Matrix Model}},}\ }%
   (\bibinfo {month} {8}\ \bibinfo {year} {2022}),\
  \Eprint{http://arxiv.org/abs/2208.14547}{arXiv:2208.14547 [hep-th]}%
  \bibAnnoteFile{NoStop}{Miller:2022fvc}%
\bibitem{Tropper:2023fjr}%
  \BibitemOpen
  \bibfield{author}{%
  \bibinfo {author} {\bibfnamefont{Adam}\ \bibnamefont{Tropper}}\ and\ \bibinfo
  {author} {\bibfnamefont{Tianli}\ \bibnamefont{Wang}},\ }%
  \bibfield{title}{%
  \enquote{\bibinfo {title} {{Lorentz symmetry and IR structure of the BFSS
  matrix model}},}\ }%
  \bibfield{journal}{%
  \Doi{10.1007/JHEP07(2023)150}{\bibinfo {journal} {JHEP}}\ }%
  \textbf{\bibinfo {volume} {07}},\ \bibinfo {pages} {150} (\bibinfo {year}
  {2023}),\ \Eprint{http://arxiv.org/abs/2303.14200}{arXiv:2303.14200
  [hep-th]}%
  \bibAnnoteFile{NoStop}{Tropper:2023fjr}%
\bibitem{Herderschee:2023bnc}%
  \BibitemOpen
  \bibfield{author}{%
  \bibinfo {author} {\bibfnamefont{Aidan}\ \bibnamefont{Herderschee}}\ and\
  \bibinfo {author} {\bibfnamefont{Juan}\ \bibnamefont{Maldacena}},\ }%
  \bibfield{title}{%
  \enquote{\bibinfo {title} {{Soft Theorems in Matrix Theory}},}\ }%
   (\bibinfo {month} {12}\ \bibinfo {year} {2023}),\
  \Eprint{http://arxiv.org/abs/2312.15111}{arXiv:2312.15111 [hep-th]}%
  \bibAnnoteFile{NoStop}{Herderschee:2023bnc}%
\bibitem{Godazgar:2019dkh}%
  \BibitemOpen
  \bibfield{author}{%
  \bibinfo {author} {\bibfnamefont{Hadi}\ \bibnamefont{Godazgar}}, \bibinfo
  {author} {\bibfnamefont{Mahdi}\ \bibnamefont{Godazgar}},\ and\ \bibinfo
  {author} {\bibfnamefont{C.~N.}\ \bibnamefont{Pope}},\ }%
  \bibfield{title}{%
  \enquote{\bibinfo {title} {{Dual gravitational charges and soft theorems}},}\
  }%
  \bibfield{journal}{%
  \Doi{10.1007/JHEP10(2019)123}{\bibinfo {journal} {JHEP}}\ }%
  \textbf{\bibinfo {volume} {10}},\ \bibinfo {pages} {123} (\bibinfo {year}
  {2019}),\ \Eprint{http://arxiv.org/abs/1908.01164}{arXiv:1908.01164
  [hep-th]}%
  \bibAnnoteFile{NoStop}{Godazgar:2019dkh}%
\bibitem{Choi:2019rlz}%
  \BibitemOpen
  \bibfield{author}{%
  \bibinfo {author} {\bibfnamefont{Sangmin}\ \bibnamefont{Choi}}\ and\ \bibinfo
  {author} {\bibfnamefont{Ratindranath}\ \bibnamefont{Akhoury}},\ }%
  \bibfield{title}{%
  \enquote{\bibinfo {title} {{Subleading soft dressings of asymptotic states in
  QED and perturbative quantum gravity}},}\ }%
  \bibfield{journal}{%
  \Doi{10.1007/JHEP09(2019)031}{\bibinfo {journal} {JHEP}}\ }%
  \textbf{\bibinfo {volume} {09}},\ \bibinfo {pages} {031} (\bibinfo {year}
  {2019}),\ \Eprint{http://arxiv.org/abs/1907.05438}{arXiv:1907.05438
  [hep-th]}%
  \bibAnnoteFile{NoStop}{Choi:2019rlz}%
\bibitem{Laddha:2017ygw}%
  \BibitemOpen
  \bibfield{author}{%
  \bibinfo {author} {\bibfnamefont{Alok}\ \bibnamefont{Laddha}}\ and\ \bibinfo
  {author} {\bibfnamefont{Ashoke}\ \bibnamefont{Sen}},\ }%
  \bibfield{title}{%
  \enquote{\bibinfo {title} {{Sub-subleading Soft Graviton Theorem in Generic
  Theories of Quantum Gravity}},}\ }%
  \bibfield{journal}{%
  \Doi{10.1007/JHEP10(2017)065}{\bibinfo {journal} {JHEP}}\ }%
  \textbf{\bibinfo {volume} {10}},\ \bibinfo {pages} {065} (\bibinfo {year}
  {2017}),\ \Eprint{http://arxiv.org/abs/1706.00759}{arXiv:1706.00759
  [hep-th]}%
  \bibAnnoteFile{NoStop}{Laddha:2017ygw}%
\bibitem{Cheung:2021yog}%
  \BibitemOpen
  \bibfield{author}{%
  \bibinfo {author} {\bibfnamefont{Clifford}\ \bibnamefont{Cheung}}, \bibinfo
  {author} {\bibfnamefont{Andreas}\ \bibnamefont{Helset}},\ and\ \bibinfo
  {author} {\bibfnamefont{Julio}\ \bibnamefont{Parra-Martinez}},\ }%
  \bibfield{title}{%
  \enquote{\bibinfo {title} {{Geometric soft theorems}},}\ }%
  \bibfield{journal}{%
  \Doi{10.1007/JHEP04(2022)011}{\bibinfo {journal} {JHEP}}\ }%
  \textbf{\bibinfo {volume} {04}},\ \bibinfo {pages} {011} (\bibinfo {year}
  {2022}),\ \Eprint{http://arxiv.org/abs/2111.03045}{arXiv:2111.03045
  [hep-th]}%
  \bibAnnoteFile{NoStop}{Cheung:2021yog}%
\bibitem{Cheung:2023qwn}%
  \BibitemOpen
  \bibfield{author}{%
  \bibinfo {author} {\bibfnamefont{Clifford}\ \bibnamefont{Cheung}}, \bibinfo
  {author} {\bibfnamefont{Maria}\ \bibnamefont{Derda}}, \bibinfo {author}
  {\bibfnamefont{Andreas}\ \bibnamefont{Helset}},\ and\ \bibinfo {author}
  {\bibfnamefont{Julio}\ \bibnamefont{Parra-Martinez}},\ }%
  \bibfield{title}{%
  \enquote{\bibinfo {title} {{Soft phonon theorems}},}\ }%
  \bibfield{journal}{%
  \Doi{10.1007/JHEP08(2023)103}{\bibinfo {journal} {JHEP}}\ }%
  \textbf{\bibinfo {volume} {08}},\ \bibinfo {pages} {103} (\bibinfo {year}
  {2023}),\ \Eprint{http://arxiv.org/abs/2301.11363}{arXiv:2301.11363
  [hep-th]}%
  \bibAnnoteFile{NoStop}{Cheung:2023qwn}%
\bibitem{Cheng:2022xyr}%
  \BibitemOpen
  \bibfield{author}{%
  \bibinfo {author} {\bibfnamefont{Peng}\ \bibnamefont{Cheng}}\ and\ \bibinfo
  {author} {\bibfnamefont{Pujian}\ \bibnamefont{Mao}},\ }%
  \bibfield{title}{%
  \enquote{\bibinfo {title} {{Soft theorems in curved spacetime}},}\ }%
  \bibfield{journal}{%
  \Doi{10.1103/PhysRevD.106.L081702}{\bibinfo {journal} {Phys. Rev. D}}\ }%
  \textbf{\bibinfo {volume} {106}},\ \bibinfo {pages} {L081702} (\bibinfo
  {year} {2022}),\ \Eprint{http://arxiv.org/abs/2206.11564}{arXiv:2206.11564
  [hep-th]}%
  \bibAnnoteFile{NoStop}{Cheng:2022xyr}%
\bibitem{Cheng:2022xgm}%
  \BibitemOpen
  \bibfield{author}{%
  \bibinfo {author} {\bibfnamefont{Peng}\ \bibnamefont{Cheng}}\ and\ \bibinfo
  {author} {\bibfnamefont{Pujian}\ \bibnamefont{Mao}},\ }%
  \bibfield{title}{%
  \enquote{\bibinfo {title} {{Soft gluon theorems in curved spacetime}},}\ }%
  \bibfield{journal}{%
  \Doi{10.1103/PhysRevD.107.065010}{\bibinfo {journal} {Phys. Rev. D}}\ }%
  \textbf{\bibinfo {volume} {107}},\ \bibinfo {pages} {065010} (\bibinfo {year}
  {2023}),\ \Eprint{http://arxiv.org/abs/2211.00031}{arXiv:2211.00031
  [hep-th]}%
  \bibAnnoteFile{NoStop}{Cheng:2022xgm}%
\bibitem{Assassi:2012zq}%
  \BibitemOpen
  \bibfield{author}{%
  \bibinfo {author} {\bibfnamefont{Valentin}\ \bibnamefont{Assassi}}, \bibinfo
  {author} {\bibfnamefont{Daniel}\ \bibnamefont{Baumann}},\ and\ \bibinfo
  {author} {\bibfnamefont{Daniel}\ \bibnamefont{Green}},\ }%
  \bibfield{title}{%
  \enquote{\bibinfo {title} {{On Soft Limits of Inflationary Correlation
  Functions}},}\ }%
  \bibfield{journal}{%
  \Doi{10.1088/1475-7516/2012/11/047}{\bibinfo {journal} {JCAP}}\ }%
  \textbf{\bibinfo {volume} {11}},\ \bibinfo {pages} {047} (\bibinfo {year}
  {2012}),\ \Eprint{http://arxiv.org/abs/1204.4207}{arXiv:1204.4207 [hep-th]}%
  \bibAnnoteFile{NoStop}{Assassi:2012zq}%
\bibitem{Derda:2024jvo}%
  \BibitemOpen
  \bibfield{author}{%
  \bibinfo {author} {\bibfnamefont{Maria}\ \bibnamefont{Derda}}, \bibinfo
  {author} {\bibfnamefont{Andreas}\ \bibnamefont{Helset}},\ and\ \bibinfo
  {author} {\bibfnamefont{Julio}\ \bibnamefont{Parra-Martinez}},\ }%
  \bibfield{title}{%
  \enquote{\bibinfo {title} {{Soft Scalars in Effective Field Theory}},}\ }%
   (\bibinfo {month} {3}\ \bibinfo {year} {2024}),\
  \Eprint{http://arxiv.org/abs/2403.12142}{arXiv:2403.12142 [hep-th]}%
  \bibAnnoteFile{NoStop}{Derda:2024jvo}%
\bibitem{Dumitrescu:2015fej}%
  \BibitemOpen
  \bibfield{author}{%
  \bibinfo {author} {\bibfnamefont{Thomas~T.}\ \bibnamefont{Dumitrescu}},
  \bibinfo {author} {\bibfnamefont{Temple}\ \bibnamefont{He}}, \bibinfo
  {author} {\bibfnamefont{Prahar}\ \bibnamefont{Mitra}},\ and\ \bibinfo
  {author} {\bibfnamefont{Andrew}\ \bibnamefont{Strominger}},\ }%
  \bibfield{title}{%
  \enquote{\bibinfo {title} {{Infinite-dimensional fermionic symmetry in
  supersymmetric gauge theories}},}\ }%
  \bibfield{journal}{%
  \Doi{10.1007/JHEP08(2021)051}{\bibinfo {journal} {JHEP}}\ }%
  \textbf{\bibinfo {volume} {08}},\ \bibinfo {pages} {051} (\bibinfo {year}
  {2021}),\ \Eprint{http://arxiv.org/abs/1511.07429}{arXiv:1511.07429
  [hep-th]}%
  \bibAnnoteFile{NoStop}{Dumitrescu:2015fej}%
\bibitem{Liu:2014vva}%
  \BibitemOpen
  \bibfield{author}{%
  \bibinfo {author} {\bibfnamefont{Zheng-Wen}\ \bibnamefont{Liu}},\ }%
  \bibfield{title}{%
  \enquote{\bibinfo {title} {{Soft theorems in maximally supersymmetric
  theories}},}\ }%
  \bibfield{journal}{%
  \Doi{10.1140/epjc/s10052-015-3304-1}{\bibinfo {journal} {Eur. Phys. J. C}}\
  }%
  \textbf{\bibinfo {volume} {75}},\ \bibinfo {pages} {105} (\bibinfo {year}
  {2015}),\ \Eprint{http://arxiv.org/abs/1410.1616}{arXiv:1410.1616 [hep-th]}%
  \bibAnnoteFile{NoStop}{Liu:2014vva}%
\bibitem{Lysov:2015jrs}%
  \BibitemOpen
  \bibfield{author}{%
  \bibinfo {author} {\bibfnamefont{Vyacheslav}\ \bibnamefont{Lysov}},\ }%
  \bibfield{title}{%
  \enquote{\bibinfo {title} {{Asymptotic Fermionic Symmetry From Soft Gravitino
  Theorem}},}\ }%
   (\bibinfo {month} {12}\ \bibinfo {year} {2015}),\
  \Eprint{http://arxiv.org/abs/1512.03015}{arXiv:1512.03015 [hep-th]}%
  \bibAnnoteFile{NoStop}{Lysov:2015jrs}%
\bibitem{Avery:2015iix}%
  \BibitemOpen
  \bibfield{author}{%
  \bibinfo {author} {\bibfnamefont{Steven~G.}\ \bibnamefont{Avery}}\ and\
  \bibinfo {author} {\bibfnamefont{Burkhard U.~W.}\ \bibnamefont{Schwab}},\ }%
  \bibfield{title}{%
  \enquote{\bibinfo {title} {{Residual Local Supersymmetry and the Soft
  Gravitino}},}\ }%
  \bibfield{journal}{%
  \Doi{10.1103/PhysRevLett.116.171601}{\bibinfo {journal} {Phys. Rev. Lett.}}\
  }%
  \textbf{\bibinfo {volume} {116}},\ \bibinfo {pages} {171601} (\bibinfo {year}
  {2016}),\ \Eprint{http://arxiv.org/abs/1512.02657}{arXiv:1512.02657
  [hep-th]}%
  \bibAnnoteFile{NoStop}{Avery:2015iix}%
\bibitem{Chen:2014xoa}%
  \BibitemOpen
  \bibfield{author}{%
  \bibinfo {author} {\bibfnamefont{Wei-Ming}\ \bibnamefont{Chen}}, \bibinfo
  {author} {\bibfnamefont{Yu-tin}\ \bibnamefont{Huang}},\ and\ \bibinfo
  {author} {\bibfnamefont{Congkao}\ \bibnamefont{Wen}},\ }%
  \bibfield{title}{%
  \enquote{\bibinfo {title} {{New Fermionic Soft Theorems for Supergravity
  Amplitudes}},}\ }%
  \bibfield{journal}{%
  \Doi{10.1103/PhysRevLett.115.021603}{\bibinfo {journal} {Phys. Rev. Lett.}}\
  }%
  \textbf{\bibinfo {volume} {115}},\ \bibinfo {pages} {021603} (\bibinfo {year}
  {2015}),\ \Eprint{http://arxiv.org/abs/1412.1809}{arXiv:1412.1809 [hep-th]}%
  \bibAnnoteFile{NoStop}{Chen:2014xoa}%
\bibitem{Bork:2015fla}%
  \BibitemOpen
  \bibfield{author}{%
  \bibinfo {author} {\bibfnamefont{L.~V.}\ \bibnamefont{Bork}}\ and\ \bibinfo
  {author} {\bibfnamefont{A.~I.}\ \bibnamefont{Onishchenko}},\ }%
  \bibfield{title}{%
  \enquote{\bibinfo {title} {{On soft theorems and form factors in $
  \mathcal{N}=4 $ SYM theory}},}\ }%
  \bibfield{journal}{%
  \Doi{10.1007/JHEP12(2015)030}{\bibinfo {journal} {JHEP}}\ }%
  \textbf{\bibinfo {volume} {12}},\ \bibinfo {pages} {030} (\bibinfo {year}
  {2015}),\ \Eprint{http://arxiv.org/abs/1506.07551}{arXiv:1506.07551
  [hep-th]}%
  \bibAnnoteFile{NoStop}{Bork:2015fla}%
\bibitem{Jain:2018fda}%
  \BibitemOpen
  \bibfield{author}{%
  \bibinfo {author} {\bibfnamefont{Diksha}\ \bibnamefont{Jain}}\ and\ \bibinfo
  {author} {\bibfnamefont{Arnab}\ \bibnamefont{Rudra}},\ }%
  \bibfield{title}{%
  \enquote{\bibinfo {title} {{Leading soft theorem for multiple gravitini}},}\
  }%
  \bibfield{journal}{%
  \Doi{10.1007/JHEP06(2019)004}{\bibinfo {journal} {JHEP}}\ }%
  \textbf{\bibinfo {volume} {06}},\ \bibinfo {pages} {004} (\bibinfo {year}
  {2019}),\ \Eprint{http://arxiv.org/abs/1811.01804}{arXiv:1811.01804
  [hep-th]}%
  \bibAnnoteFile{NoStop}{Jain:2018fda}%
\bibitem{Rao:2014zaa}%
  \BibitemOpen
  \bibfield{author}{%
  \bibinfo {author} {\bibfnamefont{Junjie}\ \bibnamefont{Rao}},\ }%
  \bibfield{title}{%
  \enquote{\bibinfo {title} {{Soft theorem of $ \mathcal{N} $ = 4 SYM in
  Grassmannian formulation}},}\ }%
  \bibfield{journal}{%
  \Doi{10.1007/JHEP02(2015)087}{\bibinfo {journal} {JHEP}}\ }%
  \textbf{\bibinfo {volume} {02}},\ \bibinfo {pages} {087} (\bibinfo {year}
  {2015}),\ \Eprint{http://arxiv.org/abs/1410.5047}{arXiv:1410.5047 [hep-th]}%
  \bibAnnoteFile{NoStop}{Rao:2014zaa}%
\bibitem{Pano:2021ewd}%
  \BibitemOpen
  \bibfield{author}{%
  \bibinfo {author} {\bibfnamefont{Yorgo}\ \bibnamefont{Pano}}, \bibinfo
  {author} {\bibfnamefont{Sabrina}\ \bibnamefont{Pasterski}},\ and\ \bibinfo
  {author} {\bibfnamefont{Andrea}\ \bibnamefont{Puhm}},\ }%
  \bibfield{title}{%
  \enquote{\bibinfo {title} {{Conformally soft fermions}},}\ }%
  \bibfield{journal}{%
  \Doi{10.1007/JHEP12(2021)166}{\bibinfo {journal} {JHEP}}\ }%
  \textbf{\bibinfo {volume} {12}},\ \bibinfo {pages} {166} (\bibinfo {year}
  {2021}),\ \Eprint{http://arxiv.org/abs/2108.11422}{arXiv:2108.11422
  [hep-th]}%
  \bibAnnoteFile{NoStop}{Pano:2021ewd}%
\bibitem{Fotopoulos:2020bqj}%
  \BibitemOpen
  \bibfield{author}{%
  \bibinfo {author} {\bibfnamefont{Angelos}\ \bibnamefont{Fotopoulos}},
  \bibinfo {author} {\bibfnamefont{Stephan}\ \bibnamefont{Stieberger}},
  \bibinfo {author} {\bibfnamefont{Tomasz~R.}\ \bibnamefont{Taylor}},\ and\
  \bibinfo {author} {\bibfnamefont{Bin}\ \bibnamefont{Zhu}},\ }%
  \bibfield{title}{%
  \enquote{\bibinfo {title} {{Extended Super BMS Algebra of Celestial CFT}},}\
  }%
  \bibfield{journal}{%
  \Doi{10.1007/JHEP09(2020)198}{\bibinfo {journal} {JHEP}}\ }%
  \textbf{\bibinfo {volume} {09}},\ \bibinfo {pages} {198} (\bibinfo {year}
  {2020}),\ \Eprint{http://arxiv.org/abs/2007.03785}{arXiv:2007.03785
  [hep-th]}%
  \bibAnnoteFile{NoStop}{Fotopoulos:2020bqj}%
\bibitem{Ball:2023qim}%
  \BibitemOpen
  \bibfield{author}{%
  \bibinfo {author} {\bibfnamefont{Adam}\ \bibnamefont{Ball}}, \bibinfo
  {author} {\bibfnamefont{Marcus}\ \bibnamefont{Spradlin}}, \bibinfo {author}
  {\bibfnamefont{Akshay}\ \bibnamefont{Yelleshpur~Srikant}},\ and\ \bibinfo
  {author} {\bibfnamefont{Anastasia}\ \bibnamefont{Volovich}},\ }%
  \bibfield{title}{%
  \enquote{\bibinfo {title} {{Supersymmetry and the Celestial Jacobi
  Identity}},}\ }%
   (\bibinfo {month} {11}\ \bibinfo {year} {2023}),\
  \Eprint{http://arxiv.org/abs/2311.01364}{arXiv:2311.01364 [hep-th]}%
  \bibAnnoteFile{NoStop}{Ball:2023qim}%
\bibitem{Bu:2021avc}%
  \BibitemOpen
  \bibfield{author}{%
  \bibinfo {author} {\bibfnamefont{Wei}\ \bibnamefont{Bu}},\ }%
  \bibfield{title}{%
  \enquote{\bibinfo {title} {{Supersymmetric celestial OPEs and soft algebras
  from the ambitwistor string worldsheet}},}\ }%
  \bibfield{journal}{%
  \Doi{10.1103/PhysRevD.105.126029}{\bibinfo {journal} {Phys. Rev. D}}\ }%
  \textbf{\bibinfo {volume} {105}},\ \bibinfo {pages} {126029} (\bibinfo {year}
  {2022}),\ \Eprint{http://arxiv.org/abs/2111.15584}{arXiv:2111.15584
  [hep-th]}%
  \bibAnnoteFile{NoStop}{Bu:2021avc}%
\bibitem{Jiang:2021ovh}%
  \BibitemOpen
  \bibfield{author}{%
  \bibinfo {author} {\bibfnamefont{Hongliang}\ \bibnamefont{Jiang}},\ }%
  \bibfield{title}{%
  \enquote{\bibinfo {title} {{Holographic chiral algebra: supersymmetry,
  infinite Ward identities, and EFTs}},}\ }%
  \bibfield{journal}{%
  \Doi{10.1007/JHEP01(2022)113}{\bibinfo {journal} {JHEP}}\ }%
  \textbf{\bibinfo {volume} {01}},\ \bibinfo {pages} {113} (\bibinfo {year}
  {2022}),\ \Eprint{http://arxiv.org/abs/2108.08799}{arXiv:2108.08799
  [hep-th]}%
  \bibAnnoteFile{NoStop}{Jiang:2021ovh}%
\bibitem{Jiang:2021xzy}%
  \BibitemOpen
  \bibfield{author}{%
  \bibinfo {author} {\bibfnamefont{Hongliang}\ \bibnamefont{Jiang}},\ }%
  \bibfield{title}{%
  \enquote{\bibinfo {title} {{Celestial superamplitude in $ \mathcal{N} $ = 4
  SYM theory}},}\ }%
  \bibfield{journal}{%
  \Doi{10.1007/JHEP08(2021)031}{\bibinfo {journal} {JHEP}}\ }%
  \textbf{\bibinfo {volume} {08}},\ \bibinfo {pages} {031} (\bibinfo {year}
  {2021}),\ \Eprint{http://arxiv.org/abs/2105.10269}{arXiv:2105.10269
  [hep-th]}%
  \bibAnnoteFile{NoStop}{Jiang:2021xzy}%
\bibitem{Banerjee:2022hpo}%
  \BibitemOpen
  \bibfield{author}{%
  \bibinfo {author} {\bibfnamefont{Nabamita}\ \bibnamefont{Banerjee}}, \bibinfo
  {author} {\bibfnamefont{Tabasum}\ \bibnamefont{Rahnuma}},\ and\ \bibinfo
  {author} {\bibfnamefont{Ranveer~Kumar}\ \bibnamefont{Singh}},\ }%
  \bibfield{title}{%
  \enquote{\bibinfo {title} {{Soft and collinear limits in $ \mathcal{N} $ = 8
  supergravity using double copy formalism}},}\ }%
  \bibfield{journal}{%
  \Doi{10.1007/JHEP04(2023)126}{\bibinfo {journal} {JHEP}}\ }%
  \textbf{\bibinfo {volume} {04}},\ \bibinfo {pages} {126} (\bibinfo {year}
  {2023}),\ \Eprint{http://arxiv.org/abs/2212.11480}{arXiv:2212.11480
  [hep-th]}%
  \bibAnnoteFile{NoStop}{Banerjee:2022hpo}%
\bibitem{Srednicki:2007qs}%
  \BibitemOpen
  \bibfield{author}{%
  \bibinfo {author} {\bibfnamefont{M.}~\bibnamefont{Srednicki}},\ }%
  \emph{\bibinfo {title} {{Quantum field theory}}}\ (\bibinfo {publisher}
  {Cambridge University Press},\ \bibinfo {year} {2007})\ ISBN \bibinfo {isbn}
  {978-0-521-86449-7, 978-0-511-26720-8}%
  \bibAnnoteFile{NoStop}{Srednicki:2007qs}%
\bibitem{Elvang:2013cua}%
  \BibitemOpen
  \bibfield{author}{%
  \bibinfo {author} {\bibfnamefont{Henriette}\ \bibnamefont{Elvang}}\ and\
  \bibinfo {author} {\bibfnamefont{Yu-tin}\ \bibnamefont{Huang}},\ }%
  \bibfield{title}{%
  \enquote{\bibinfo {title} {{Scattering Amplitudes}},}\ }%
   (\bibinfo {month} {8}\ \bibinfo {year} {2013}),\
  \Eprint{http://arxiv.org/abs/1308.1697}{arXiv:1308.1697 [hep-th]}%
  \bibAnnoteFile{NoStop}{Elvang:2013cua}%
\bibitem{Krishna:2023fxg}%
  \BibitemOpen
  \bibfield{author}{%
  \bibinfo {author} {\bibfnamefont{Hare}\ \bibnamefont{Krishna}}\ and\ \bibinfo
  {author} {\bibfnamefont{Biswajit}\ \bibnamefont{Sahoo}},\ }%
  \bibfield{title}{%
  \enquote{\bibinfo {title} {{Universality of Loop Corrected Soft Theorems in
  4d}},}\ }%
   (\bibinfo {month} {8}\ \bibinfo {year} {2023}),\
  \Eprint{http://arxiv.org/abs/2308.16807}{arXiv:2308.16807 [hep-th]}%
  \bibAnnoteFile{NoStop}{Krishna:2023fxg}%
\bibitem{Laddha:2018myi}%
  \BibitemOpen
  \bibfield{author}{%
  \bibinfo {author} {\bibfnamefont{Alok}\ \bibnamefont{Laddha}}\ and\ \bibinfo
  {author} {\bibfnamefont{Ashoke}\ \bibnamefont{Sen}},\ }%
  \bibfield{title}{%
  \enquote{\bibinfo {title} {{Logarithmic Terms in the Soft Expansion in Four
  Dimensions}},}\ }%
  \bibfield{journal}{%
  \Doi{10.1007/JHEP10(2018)056}{\bibinfo {journal} {JHEP}}\ }%
  \textbf{\bibinfo {volume} {10}},\ \bibinfo {pages} {056} (\bibinfo {year}
  {2018}),\ \Eprint{http://arxiv.org/abs/1804.09193}{arXiv:1804.09193
  [hep-th]}%
  \bibAnnoteFile{NoStop}{Laddha:2018myi}%
\bibitem{Sahoo:2018lxl}%
  \BibitemOpen
  \bibfield{author}{%
  \bibinfo {author} {\bibfnamefont{Biswajit}\ \bibnamefont{Sahoo}}\ and\
  \bibinfo {author} {\bibfnamefont{Ashoke}\ \bibnamefont{Sen}},\ }%
  \bibfield{title}{%
  \enquote{\bibinfo {title} {{Classical and Quantum Results on Logarithmic
  Terms in the Soft Theorem in Four Dimensions}},}\ }%
  \bibfield{journal}{%
  \Doi{10.1007/JHEP02(2019)086}{\bibinfo {journal} {JHEP}}\ }%
  \textbf{\bibinfo {volume} {02}},\ \bibinfo {pages} {086} (\bibinfo {year}
  {2019}),\ \Eprint{http://arxiv.org/abs/1808.03288}{arXiv:1808.03288
  [hep-th]}%
  \bibAnnoteFile{NoStop}{Sahoo:2018lxl}%
\bibitem{Addazi:2019mjh}%
  \BibitemOpen
  \bibfield{author}{%
  \bibinfo {author} {\bibfnamefont{Andrea}\ \bibnamefont{Addazi}}, \bibinfo
  {author} {\bibfnamefont{Massimo}\ \bibnamefont{Bianchi}},\ and\ \bibinfo
  {author} {\bibfnamefont{Gabriele}\ \bibnamefont{Veneziano}},\ }%
  \bibfield{title}{%
  \enquote{\bibinfo {title} {{Soft gravitational radiation from
  ultra-relativistic collisions at sub- and sub-sub-leading order}},}\ }%
  \bibfield{journal}{%
  \Doi{10.1007/JHEP05(2019)050}{\bibinfo {journal} {JHEP}}\ }%
  \textbf{\bibinfo {volume} {05}},\ \bibinfo {pages} {050} (\bibinfo {year}
  {2019}),\ \Eprint{http://arxiv.org/abs/1901.10986}{arXiv:1901.10986
  [hep-th]}%
  \bibAnnoteFile{NoStop}{Addazi:2019mjh}%
\bibitem{Agrawal:2023zea}%
  \BibitemOpen
  \bibfield{author}{%
  \bibinfo {author} {\bibfnamefont{Shreyansh}\ \bibnamefont{Agrawal}}, \bibinfo
  {author} {\bibfnamefont{Laura}\ \bibnamefont{Donnay}}, \bibinfo {author}
  {\bibfnamefont{Kevin}\ \bibnamefont{Nguyen}},\ and\ \bibinfo {author}
  {\bibfnamefont{Romain}\ \bibnamefont{Ruzziconi}},\ }%
  \bibfield{title}{%
  \enquote{\bibinfo {title} {{Logarithmic soft graviton theorems from
  superrotation Ward identities}},}\ }%
   (\bibinfo {month} {9}\ \bibinfo {year} {2023}),\
  \Eprint{http://arxiv.org/abs/2309.11220}{arXiv:2309.11220 [hep-th]}%
  \bibAnnoteFile{NoStop}{Agrawal:2023zea}%
\bibitem{Ciafaloni:2018uwe}%
  \BibitemOpen
  \bibfield{author}{%
  \bibinfo {author} {\bibfnamefont{Marcello}\ \bibnamefont{Ciafaloni}},
  \bibinfo {author} {\bibfnamefont{Dimitri}\ \bibnamefont{Colferai}},\ and\
  \bibinfo {author} {\bibfnamefont{Gabriele}\ \bibnamefont{Veneziano}},\ }%
  \bibfield{title}{%
  \enquote{\bibinfo {title} {{Infrared features of gravitational scattering and
  radiation in the eikonal approach}},}\ }%
  \bibfield{journal}{%
  \Doi{10.1103/PhysRevD.99.066008}{\bibinfo {journal} {Phys. Rev. D}}\ }%
  \textbf{\bibinfo {volume} {99}},\ \bibinfo {pages} {066008} (\bibinfo {year}
  {2019}),\ \Eprint{http://arxiv.org/abs/1812.08137}{arXiv:1812.08137
  [hep-th]}%
  \bibAnnoteFile{NoStop}{Ciafaloni:2018uwe}%
\bibitem{Mao:2023rca}%
  \BibitemOpen
  \bibfield{author}{%
  \bibinfo {author} {\bibfnamefont{Pujian}\ \bibnamefont{Mao}}\ and\ \bibinfo
  {author} {\bibfnamefont{Kai-Yu}\ \bibnamefont{Zhang}},\ }%
  \bibfield{title}{%
  \enquote{\bibinfo {title} {{Soft theorems in de Sitter spacetime}},}\ }%
   (\bibinfo {month} {8}\ \bibinfo {year} {2023}),\
  \Eprint{http://arxiv.org/abs/2308.08861}{arXiv:2308.08861 [hep-th]}%
  \bibAnnoteFile{NoStop}{Mao:2023rca}%
\bibitem{Elvang:2016qvq}%
  \BibitemOpen
  \bibfield{author}{%
  \bibinfo {author} {\bibfnamefont{Henriette}\ \bibnamefont{Elvang}}, \bibinfo
  {author} {\bibfnamefont{Callum R.~T.}\ \bibnamefont{Jones}},\ and\ \bibinfo
  {author} {\bibfnamefont{Stephen~G.}\ \bibnamefont{Naculich}},\ }%
  \bibfield{title}{%
  \enquote{\bibinfo {title} {{Soft Photon and Graviton Theorems in Effective
  Field Theory}},}\ }%
  \bibfield{journal}{%
  \Doi{10.1103/PhysRevLett.118.231601}{\bibinfo {journal} {Phys. Rev. Lett.}}\
  }%
  \textbf{\bibinfo {volume} {118}},\ \bibinfo {pages} {231601} (\bibinfo {year}
  {2017}),\ \Eprint{http://arxiv.org/abs/1611.07534}{arXiv:1611.07534
  [hep-th]}%
  \bibAnnoteFile{NoStop}{Elvang:2016qvq}%
\bibitem{Laddha:2017vfh}%
  \BibitemOpen
  \bibfield{author}{%
  \bibinfo {author} {\bibfnamefont{Alok}\ \bibnamefont{Laddha}}\ and\ \bibinfo
  {author} {\bibfnamefont{Prahar}\ \bibnamefont{Mitra}},\ }%
  \bibfield{title}{%
  \enquote{\bibinfo {title} {{Asymptotic Symmetries and Subleading Soft Photon
  Theorem in Effective Field Theories}},}\ }%
  \bibfield{journal}{%
  \Doi{10.1007/JHEP05(2018)132}{\bibinfo {journal} {JHEP}}\ }%
  \textbf{\bibinfo {volume} {05}},\ \bibinfo {pages} {132} (\bibinfo {year}
  {2018}),\ \Eprint{http://arxiv.org/abs/1709.03850}{arXiv:1709.03850
  [hep-th]}%
  \bibAnnoteFile{NoStop}{Laddha:2017vfh}%
\bibitem{Awada:1985by}%
  \BibitemOpen
  \bibfield{author}{%
  \bibinfo {author} {\bibfnamefont{M.~A.}\ \bibnamefont{Awada}}, \bibinfo
  {author} {\bibfnamefont{G.~W.}\ \bibnamefont{Gibbons}},\ and\ \bibinfo
  {author} {\bibfnamefont{W.~T.}\ \bibnamefont{Shaw}},\ }%
  \bibfield{title}{%
  \enquote{\bibinfo {title} {{CONFORMAL SUPERGRAVITY, TWISTORS AND THE SUPER
  BMS GROUP}},}\ }%
  \bibfield{journal}{%
  \Doi{10.1016/S0003-4916(86)80023-9}{\bibinfo {journal} {Annals Phys.}}\ }%
  \textbf{\bibinfo {volume} {171}},\ \bibinfo {pages} {52} (\bibinfo {year}
  {1986})%
  \bibAnnoteFile{NoStop}{Awada:1985by}%
\bibitem{Henneaux:2020ekh}%
  \BibitemOpen
  \bibfield{author}{%
  \bibinfo {author} {\bibfnamefont{Marc}\ \bibnamefont{Henneaux}}, \bibinfo
  {author} {\bibfnamefont{Javier}\ \bibnamefont{Matulich}},\ and\ \bibinfo
  {author} {\bibfnamefont{Turmoli}\ \bibnamefont{Neogi}},\ }%
  \bibfield{title}{%
  \enquote{\bibinfo {title} {{Asymptotic realization of the super-BMS algebra
  at spatial infinity}},}\ }%
  \bibfield{journal}{%
  \Doi{10.1103/PhysRevD.101.126016}{\bibinfo {journal} {Phys. Rev. D}}\ }%
  \textbf{\bibinfo {volume} {101}},\ \bibinfo {pages} {126016} (\bibinfo {year}
  {2020}),\ \Eprint{http://arxiv.org/abs/2004.07299}{arXiv:2004.07299
  [hep-th]}%
  \bibAnnoteFile{NoStop}{Henneaux:2020ekh}%
\bibitem{Fuentealba:2021xhn}%
  \BibitemOpen
  \bibfield{author}{%
  \bibinfo {author} {\bibfnamefont{Oscar}\ \bibnamefont{Fuentealba}}, \bibinfo
  {author} {\bibfnamefont{Marc}\ \bibnamefont{Henneaux}}, \bibinfo {author}
  {\bibfnamefont{Sucheta}\ \bibnamefont{Majumdar}}, \bibinfo {author}
  {\bibfnamefont{Javier}\ \bibnamefont{Matulich}},\ and\ \bibinfo {author}
  {\bibfnamefont{Turmoli}\ \bibnamefont{Neogi}},\ }%
  \bibfield{title}{%
  \enquote{\bibinfo {title} {{Local supersymmetry and the square roots of
  Bondi-Metzner-Sachs supertranslations}},}\ }%
  \bibfield{journal}{%
  \Doi{10.1103/PhysRevD.104.L121702}{\bibinfo {journal} {Phys. Rev. D}}\ }%
  \textbf{\bibinfo {volume} {104}},\ \bibinfo {pages} {L121702} (\bibinfo
  {year} {2021}),\ \Eprint{http://arxiv.org/abs/2108.07825}{arXiv:2108.07825
  [hep-th]}%
  \bibAnnoteFile{NoStop}{Fuentealba:2021xhn}%
\bibitem{Banerjee:2022abf}%
  \BibitemOpen
  \bibfield{author}{%
  \bibinfo {author} {\bibfnamefont{Nabamita}\ \bibnamefont{Banerjee}}, \bibinfo
  {author} {\bibfnamefont{Arpita}\ \bibnamefont{Mitra}}, \bibinfo {author}
  {\bibfnamefont{Debangshu}\ \bibnamefont{Mukherjee}},\ and\ \bibinfo {author}
  {\bibfnamefont{H.~R.}\ \bibnamefont{Safari}},\ }%
  \bibfield{title}{%
  \enquote{\bibinfo {title} {{Supersymmetrization of deformed BMS algebras}},}\
  }%
  \bibfield{journal}{%
  \Doi{10.1140/epjc/s10052-022-11036-y}{\bibinfo {journal} {Eur. Phys. J. C}}\
  }%
  \textbf{\bibinfo {volume} {83}},\ \bibinfo {pages} {3} (\bibinfo {year}
  {2023}),\ \Eprint{http://arxiv.org/abs/2201.09853}{arXiv:2201.09853
  [hep-th]}%
  \bibAnnoteFile{NoStop}{Banerjee:2022abf}%
\bibitem{Boulanger:2023gpw}%
  \BibitemOpen
  \bibfield{author}{%
  \bibinfo {author} {\bibfnamefont{Nicolas}\ \bibnamefont{Boulanger}}, \bibinfo
  {author} {\bibfnamefont{Yannick}\ \bibnamefont{Herfray}},\ and\ \bibinfo
  {author} {\bibfnamefont{No\'emie}\ \bibnamefont{Parrini}},\ }%
  \bibfield{title}{%
  \enquote{\bibinfo {title} {{Conformal boundaries of Minkowski superspace and
  their super cuts}},}\ }%
  \bibfield{journal}{%
  \Doi{10.1007/JHEP02(2024)177}{\bibinfo {journal} {JHEP}}\ }%
  \textbf{\bibinfo {volume} {02}},\ \bibinfo {pages} {177} (\bibinfo {year}
  {2024}),\ \Eprint{http://arxiv.org/abs/2312.11222}{arXiv:2312.11222
  [hep-th]}%
  \bibAnnoteFile{NoStop}{Boulanger:2023gpw}%
\bibitem{Banerjee:2022lnz}%
  \BibitemOpen
  \bibfield{author}{%
  \bibinfo {author} {\bibfnamefont{Nabamita}\ \bibnamefont{Banerjee}}, \bibinfo
  {author} {\bibfnamefont{Tabasum}\ \bibnamefont{Rahnuma}},\ and\ \bibinfo
  {author} {\bibfnamefont{Ranveer~Kumar}\ \bibnamefont{Singh}},\ }%
  \bibfield{title}{%
  \enquote{\bibinfo {title} {{Asymptotic symmetry algebra of N=8
  supergravity}},}\ }%
  \bibfield{journal}{%
  \Doi{10.1103/PhysRevD.109.046010}{\bibinfo {journal} {Phys. Rev. D}}\ }%
  \textbf{\bibinfo {volume} {109}},\ \bibinfo {pages} {046010} (\bibinfo {year}
  {2024}),\ \Eprint{http://arxiv.org/abs/2212.12133}{arXiv:2212.12133
  [hep-th]}%
  \bibAnnoteFile{NoStop}{Banerjee:2022lnz}%
\bibitem{Freedman:2012zz}%
  \BibitemOpen
  \bibfield{author}{%
  \bibinfo {author} {\bibfnamefont{Daniel~Z.}\ \bibnamefont{Freedman}}\ and\
  \bibinfo {author} {\bibfnamefont{Antoine}\ \bibnamefont{Van~Proeyen}},\ }%
  \emph{\bibinfo {title} {{Supergravity}}}\ (\bibinfo {publisher} {Cambridge
  Univ. Press},\ \bibinfo {address} {Cambridge, UK},\ \bibinfo {year} {2012})\
  ISBN \bibinfo {isbn} {978-1-139-36806-3, 978-0-521-19401-3}%
  \bibAnnoteFile{NoStop}{Freedman:2012zz}%
\bibitem{Tropper:2024evi}%
  \BibitemOpen
  \bibfield{author}{%
  \bibinfo {author} {\bibfnamefont{Adam}\ \bibnamefont{Tropper}},\ }%
  \bibfield{title}{%
  \enquote{\bibinfo {title} {{Symmetries of the Celestial Supersphere}},}\ }%
   (\bibinfo {month} {12}\ \bibinfo {year} {2024}),\
  \Eprint{http://arxiv.org/abs/2412.13113}{arXiv:2412.13113 [hep-th]}%
  \bibAnnoteFile{NoStop}{Tropper:2024evi}%
\bibitem{Kapec:2022axw}%
  \BibitemOpen
  \bibfield{author}{%
  \bibinfo {author} {\bibfnamefont{Daniel}\ \bibnamefont{Kapec}}, \bibinfo
  {author} {\bibfnamefont{Y.~T.~Albert}\ \bibnamefont{Law}},\ and\ \bibinfo
  {author} {\bibfnamefont{Sruthi~A.}\ \bibnamefont{Narayanan}},\ }%
  \bibfield{title}{%
  \enquote{\bibinfo {title} {{Soft scalars and the geometry of the space of
  celestial conformal field theories}},}\ }%
  \bibfield{journal}{%
  \Doi{10.1103/PhysRevD.107.046024}{\bibinfo {journal} {Phys. Rev. D}}\ }%
  \textbf{\bibinfo {volume} {107}},\ \bibinfo {pages} {046024} (\bibinfo {year}
  {2023}),\ \Eprint{http://arxiv.org/abs/2205.10935}{arXiv:2205.10935
  [hep-th]}%
  \bibAnnoteFile{NoStop}{Kapec:2022axw}%
\bibitem{WIP:Seiberg-Witten}%
  \BibitemOpen
  \bibfield{author}{%
  \bibinfo {author} {\bibfnamefont{Erin}\ \bibnamefont{Crawley}}, \bibinfo
  {author} {\bibfnamefont{Andrew}\ \bibnamefont{Strominger}},\ and\ \bibinfo
  {author} {\bibfnamefont{Adam}\ \bibnamefont{Tropper}},\ }%
  \bibfield{title}{%
  \enquote{\bibinfo {title} {{Chiral Soft Algebras for $\mathcal{N} = 2$ Gauge
  Theory}},}\ }%
   (\bibinfo {month} {7}\ \bibinfo {year} {2024}),\
  \Eprint{http://arxiv.org/abs/2407.16752}{arXiv:2407.16752 [hep-th]}%
  \bibAnnoteFile{NoStop}{WIP:Seiberg-Witten}%
\bibitem{Nair:1988bq}%
  \BibitemOpen
  \bibfield{author}{%
  \bibinfo {author} {\bibfnamefont{V.~P.}\ \bibnamefont{Nair}},\ }%
  \bibfield{title}{%
  \enquote{\bibinfo {title} {{A Current Algebra for Some Gauge Theory
  Amplitudes}},}\ }%
  \bibfield{journal}{%
  \Doi{10.1016/0370-2693(88)91471-2}{\bibinfo {journal} {Phys. Lett. B}}\ }%
  \textbf{\bibinfo {volume} {214}},\ \bibinfo {pages} {215--218} (\bibinfo
  {year} {1988})%
  \bibAnnoteFile{NoStop}{Nair:1988bq}%
\bibitem{Elvang:2011fx}%
  \BibitemOpen
  \bibfield{author}{%
  \bibinfo {author} {\bibfnamefont{Henriette}\ \bibnamefont{Elvang}}, \bibinfo
  {author} {\bibfnamefont{Yu-tin}\ \bibnamefont{Huang}},\ and\ \bibinfo
  {author} {\bibfnamefont{Cheng}\ \bibnamefont{Peng}},\ }%
  \bibfield{title}{%
  \enquote{\bibinfo {title} {{On-shell superamplitudes in N\ensuremath{<}4
  SYM}},}\ }%
  \bibfield{journal}{%
  \Doi{10.1007/JHEP09(2011)031}{\bibinfo {journal} {JHEP}}\ }%
  \textbf{\bibinfo {volume} {09}},\ \bibinfo {pages} {031} (\bibinfo {year}
  {2011}),\ \Eprint{http://arxiv.org/abs/1102.4843}{arXiv:1102.4843 [hep-th]}%
  \bibAnnoteFile{NoStop}{Elvang:2011fx}%
\bibitem{He:2014bga}%
  \BibitemOpen
  \bibfield{author}{%
  \bibinfo {author} {\bibfnamefont{Song}\ \bibnamefont{He}}, \bibinfo {author}
  {\bibfnamefont{Yu-tin}\ \bibnamefont{Huang}},\ and\ \bibinfo {author}
  {\bibfnamefont{Congkao}\ \bibnamefont{Wen}},\ }%
  \bibfield{title}{%
  \enquote{\bibinfo {title} {{Loop Corrections to Soft Theorems in Gauge
  Theories and Gravity}},}\ }%
  \bibfield{journal}{%
  \Doi{10.1007/JHEP12(2014)115}{\bibinfo {journal} {JHEP}}\ }%
  \textbf{\bibinfo {volume} {12}},\ \bibinfo {pages} {115} (\bibinfo {year}
  {2014}),\ \Eprint{http://arxiv.org/abs/1405.1410}{arXiv:1405.1410 [hep-th]}%
  \bibAnnoteFile{NoStop}{He:2014bga}%
\bibitem{Guevara:2019ypd}%
  \BibitemOpen
  \bibfield{author}{%
  \bibinfo {author} {\bibfnamefont{Alfredo}\ \bibnamefont{Guevara}},\ }%
  \bibfield{title}{%
  \enquote{\bibinfo {title} {{Notes on Conformal Soft Theorems and Recursion
  Relations in Gravity}},}\ }%
   (\bibinfo {month} {6}\ \bibinfo {year} {2019}),\
  \Eprint{http://arxiv.org/abs/1906.07810}{arXiv:1906.07810 [hep-th]}%
  \bibAnnoteFile{NoStop}{Guevara:2019ypd}%
\end{thebibliography}%
\bibliographystyle{apsrev4-1long}

\end{document}